\documentclass
[aps,prl,amsfonts,amssymb,twocolumn,amsmath,preprintnumbers,nofootinbib,floatfix,superscriptaddress,longbibliography]{revtex4-2}%
\usepackage[dvips]{graphics}
\usepackage{graphicx}
\usepackage{bm}
\usepackage{amsmath}
\usepackage{amsfonts}
\usepackage{amssymb}
\usepackage{xcolor}
\usepackage{subfigure}
\usepackage{tabularx}
\usepackage{multirow}
\usepackage{braket}
\usepackage{commath}
\usepackage[colorlinks,linkcolor=blue,anchorcolor=blue,citecolor=blue,urlcolor=blue]%
{hyperref}%
\providecommand{\U}[1]{\protect\rule{.1in}{.1in}}
\begin{document}

%\title{Intrinsic dipole Hall effect as non-contact probe of topological transition\\ between magnetically insulating phases in twisted MoTe$_2$ bilayer}

%\title{dipole Hall effect of magnetically insulating phases in twisted MoTe$_2$ and non-contact probe of topological transition}

%\title{Dipole Hall effect of magnetic insulators and contact-free signature of topological transitions in tMoTe$_2$ moir\'{e}}

\title{Intrinsic Dipole Hall effect in twisted MoTe$_2$: magnetoelectricity and contact-free signatures of topological transitions}

\author{Feng-Ren Fan}
\thanks{These authors contributed equally to this work.}
\affiliation{New Cornerstone Science Laboratory, Department of Physics, University of Hong Kong, Hong Kong, China}
\affiliation{HKU-UCAS Joint Institute of Theoretical and Computational Physics at Hong Kong, China}
\author{Cong Xiao}
\thanks{These authors contributed equally to this work.}
\affiliation{Institute of Applied Physics and Materials Engineering, University of Macau, Taipa, Macau, China}
\affiliation{HKU-UCAS Joint Institute of Theoretical and Computational Physics at Hong Kong, China}
\author{Wang Yao}
\email[]{wangyao@hku.hk}
\affiliation{New Cornerstone Science Laboratory, Department of Physics, University of Hong Kong, Hong Kong, China}
\affiliation{HKU-UCAS Joint Institute of Theoretical and Computational Physics at Hong Kong, China}

\begin{abstract}
We discover an intrinsic dipole Hall effect in a variety of magnetic insulating states at integer fillings of twisted MoTe$_2$ moir\'e superlattice, including topologically trivial and nontrivial ferro-, antiferro-, and ferri-magnetic configurations. The dipole Hall current, in linear response to in-plane electric field, generates an in-plane orbital magnetization $M_{\parallel}$ along the field, through which an AC field can drive magnetization oscillation up to THz range. Upon the continuous topological phase transitions from trivial to quantum anomalous Hall states in both ferromagnetic and antiferromagnetic configurations, the dipole Hall current and $M_{\parallel}$ have an abrupt sign change, enabling contact free detection of the transitions through the magnetic stray field. In configurations where the linear response is forbidden by symmetry, the dipole Hall current and $M_{\parallel}$ appear as a crossed nonlinear response to both in-plane and out-of-plane electric fields. These magnetoelectric phenomena showcase novel functionalities of insulators from the interplay between magnetism, topology and electrical polarization. 
\end{abstract}

\maketitle

\emph{{\color{blue} Introduction.}}--
In magnetic insulators, in-depth exploration of magnon transport has unveiled promising opportunities for low-power-consumption information technologies~\cite{chumak2015Magnon, yu2018Magnon}. As the electrical counterpart of magnon, a charge neutral elementary excitation carrying electric dipole can also transport energy, momentum, as well as electrical polarization in an insulator~\cite{bauer2023Polarization}. Such a concept has caught limited attentions until very recently in the context of  ferroelectrics where the dipole excitation corresponds to certain phonon~\cite{shen2022Magnonferron, chotorlishvili2013Dynamics, bauer2021Theory, chen2021Narrow, li2021Subterahertz, tang2022Thermoelectric, bauer2023Polarization}. Notably, flow of electric dipole in directions perpendicular to its moment is detectable via the accompanied magnetic stray field~\cite{tang2022Thermoelectric}.

\begin{figure}[phtb]
\includegraphics[width=0.48\textwidth]{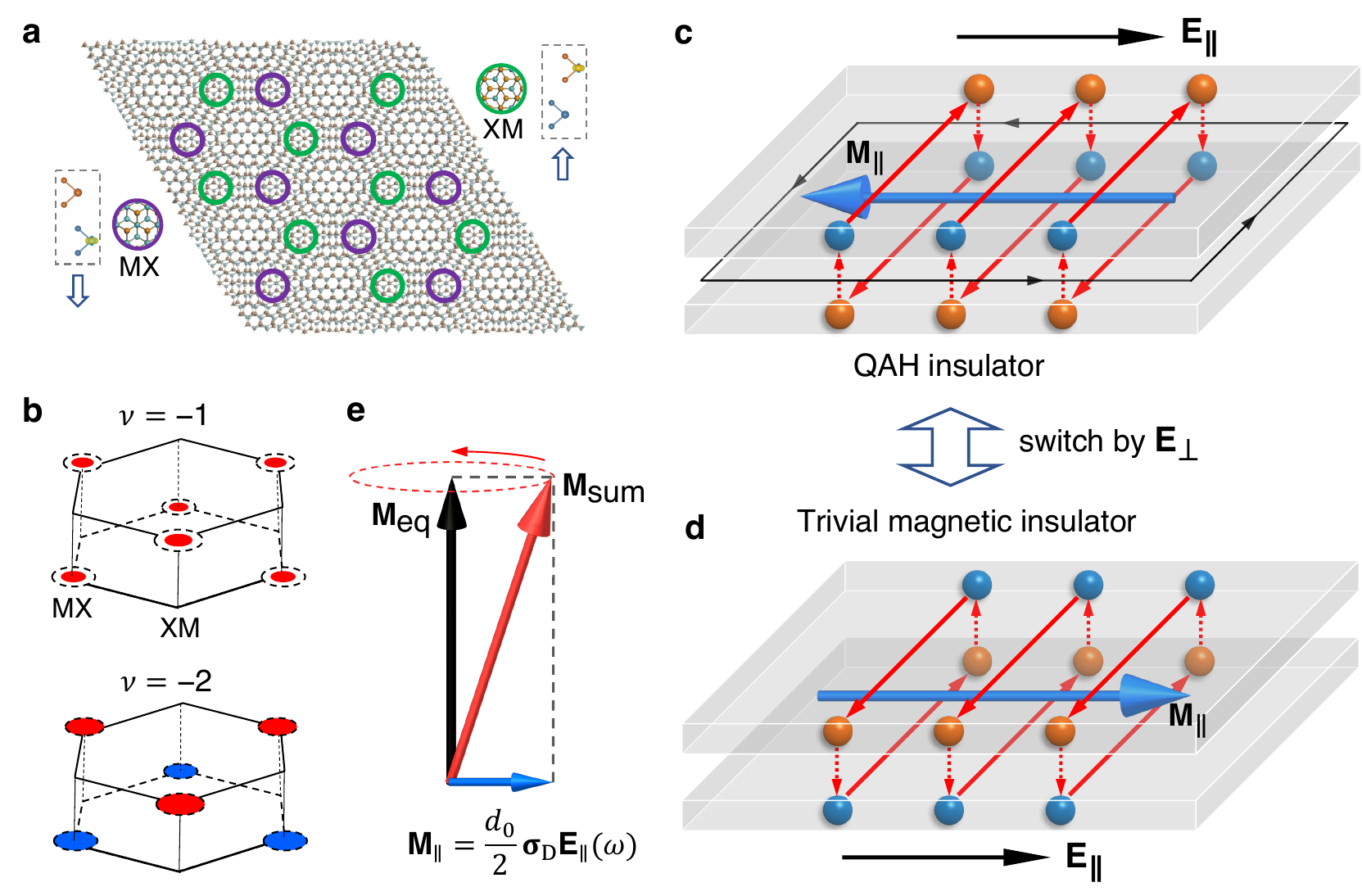}
\caption{Schematic of the dipole Hall effect in twisted MoTe$_2$.
(a) Schematic of twisted R-stack TMDs bilayer, featuring opposite out-of-plane electrical polarizations at MX and XM stacking regions. (b) FM and AFM insulating states at filling $\nu = -1$ and $-2$. Dashed circles denote the moiré orbitals of doped carriers, polarized in opposite layers at MX and XM sites. The colored area signifies both the filling fraction and spin. 
(c)-(d) Dipole Hall response to in-plane electric field $E_{\parallel}$ for (c) QAH and (d) trivial magnetic insulating state.
The red dashed arrows denote the interlayer tunneling current due to annihilation of accumulated dipole on edges, and the red solid arrows denote the layer counter flows corresponding to the bulk dipole Hall current. Together, they form a current loop that generates the in-plane magnetization in bulk.
Upon the topological phase transition controlled by perpendicular field $E_{\perp}$, the dipole Hall current and associated orbital magnetization $M_{\parallel}$ (blue arrows) have an abrupt sign change.
(e) Equilibrium (black) and total (red) magnetization in an AC field $E_{\parallel}(\omega)$.
}%
\label{fig:schematic}%
\end{figure}

Rhombohedral (R) homobilayers of transition metal dichalcogenides (TMDs) is a versatile platform that hosts ferroelectricity~\cite{weston2022Interfacial, wang2022Interfacial}, intrinsic magnetism~\cite{anderson2023Programming}, and nontrivial topology~\cite{cai2023Signatures, park2023Observation, xu2023Observation, zeng2023Thermodynamica}. With inversion symmetry broken, the electron affinity difference between atoms leads to an interlayer electrical polarization having opposite signs at the MX and XM stacking configurations that are related by an interlayer sliding [Fig.~\ref{fig:schematic}(a)]~\cite{li2017Binary, zhao2021Universal, liang2022Optically, rogee2022Ferroelectricity, weston2022Interfacial, wang2022Interfacial}. In commensurate and marginally twisted homobilayers, switching of the polarization accompanied by sliding can be achieved by a sizable out-of-plane electric field~\cite{rogee2022Ferroelectricity, weston2022Interfacial, wang2022Interfacial, viznerstern2021Interfacial, woods2021Chargepolarized, yasuda2021Stackingengineered}, forming the basis of ferroelectric functionalities. 

Twisting the homobilayers by a modest angle results in a moiré pattern where stacking registries alternate between MX and XM with few nm periodicity [Fig.~\ref{fig:schematic}(a)]. The stacking dependent electrical polarization then becomes an antiferroelectric background pinned by the moiré, in which doped carriers experience a hexagonal superlattice with two degenerate moiré orbitals at MX and XM regions polarized in opposite layers respectively~\cite{wu2019Topological, yu2020Giant}. Berry phase from such layer texture of carrier manifests as an emergent magnetic field of quantized flux per moiré cell~\cite{yu2020Giant, zhai2020Theory}, which underlies nontrivial topology of low-energy minibands~\cite{wu2019Topological}. With the intrinsic ferromagnetism from Coulomb exchange between moiré orbitals~\cite{anderson2023Programming}, this system has become an exciting platform for exploring quantum anomalous Hall (QAH) effects, where both integer and fractional QAH effects were observed in twisted MoTe$_2$~\cite{cai2023Signatures, park2023Observation, xu2023Observation, zeng2023Thermodynamica}. Through the antiferroelectric nature of the carrier wavefunction, both the magnetism and topology can be manipulated by a modest perpendicular electric field ($E_{\perp}$)~\cite{yu2020Giant, anderson2023Programming}. At filling factor $\nu=-1$, i.e. one hole per moiré cell, experiment has reported a continuous topological phase transition from ferromagnetic (FM) QAH to a trivial FM, with the increase of $E_{\perp}$~\cite{park2023Observation}. Hartree-Fock calculations further suggest the existence of an antiferromagnetic (AFM) state at $\nu=-2$, featuring a topological transition from trivial to an AFM QAH upon the increase of $E_{\perp}$~\cite{fan2024Orbital, jiang2018Antiferromagnetic, liu2023Gatetunable}.

%
% figure-1: schematic of dipole Hall in tMoTe2

Here we discover an intrinsic dipole Hall effect generally present in a variety of magnetic insulating states at integer filling factors in twisted R stack TMDs homobilayers. In the insulating bulk, a pure flow of interlayer dipole excitation of the doped carrier is generated due to the {\it dipole Berry curvature} in superlattice minibands. Such quantum geometric origin allows linear dipole current response to an in-plane electric field $E_{\parallel}$, rather than the usual field gradient. On top of the equilibrium magnetization out-of-plane, the dipole Hall current corresponds to an in-plane orbital magnetization $M_{\parallel}$ along $E_{\parallel}$. Through this magnetoelectric response, an AC electrical field can thus drive magnetization oscillations up to the terahertz range.
Remarkably, upon the continuous topological quantum phase transitions tuned by $E_{\perp}$ in both the $\nu=-1$ FM and $\nu=-2$ AFM states, the dipole Hall conductivity and the associated $M_{\parallel}$ have an abrupt change, enabling contact-free detection of the transitions through the magnetic stray field. In the $\nu=-1$ ferro- and a $\nu=-3$ ferri-magnetic configurations where this linear response is forbidden at $E_{\perp}=0$ by the $C_{2y}T$ symmetry, the dipole Hall current and $M_{\parallel}$ appear as a nonlinear response to both $E_{\perp}$ and $E_{\parallel}$.

\emph{{\color{blue} Quantum geometric origin.}}--
For electrons in a coupled bilayer, the dipole current operator reads
$\hat{j}_{a}=\frac{e}{2}\{\hat{v}_{a},\hat{\sigma}_{z}\}$. Here $\hat{\sigma}_{z}$ is the Pauli matrix in the layer index subspace, and represents the interlayer charge dipole $\boldsymbol{\hat{p}%
}=ed_{0}\hat{\sigma}_{z}\boldsymbol{\hat{z}}$, where $d_{0}$ is the interlayer distance.
By the semiclassical theory, the dipole current contributed by an electron is given by (see Supplemental Material \cite{supp}):
\begin{equation}
J_{a}^{n}(\boldsymbol{k})=j_{a}^{n}
(\boldsymbol{k})+\Upsilon_{ab}^{n}(\boldsymbol{k})E_{\parallel,b},
\label{Abelian}%
\end{equation}
where $n$ and $\boldsymbol{k}$ are the band index and the wave vector, respectively, and summation over repeated Cartesian indices $a,b$ is implied.
The first term on the right side is the expectation value of $\hat{j}_{a}$. The second term is the {\it anomalous dipole current} induced by field, where
\begin{equation}
\Upsilon_{ab}^{n}(\boldsymbol{k})=2e\hbar\operatorname{Im}\sum_{n^{\prime}\neq
n}\frac{j_{a}^{nn^{\prime}}(\boldsymbol{k})v_{b}^{n^{\prime}n}(\boldsymbol{k}%
)}{[\varepsilon_{n}(\boldsymbol{k})-\varepsilon_{n^{\prime}}(\boldsymbol{k}%
)]^{2}} \label{geometric}%
\end{equation}
can be termed, in the same spirit of the spin Berry curvature \cite{Sun2016}, as {\it dipole Berry curvature}. Here $\varepsilon_{n}(\boldsymbol{k})$ is the band energy, and the
numerator involves the interband matrix elements of interlayer-dipole current and
velocity operators. 

Summing over $\boldsymbol{k}$ in the filled bands of the insulator yields the total dipole current density. We find an intrinsic dipole current in linear response to $E_{\parallel}$: $j_{\mathrm{D},a}=\sigma_{ab}E_{\parallel,b}$, where
$\sigma_{ab}=\sum_{n}\int[d\boldsymbol{k}]f_{0}(\boldsymbol{k})\Upsilon_{ab}%
^{n}(\boldsymbol{k})
$, $f_{0}$ the equilibrium Fermi distribution,
and $[d\boldsymbol{k}]$ is shorthand for $
d\boldsymbol{k}/(2\pi)^{2}$.
For the off-diagonal component of dipole conductivity $\sigma_{ab}$, the threefold rotation
symmetry $C_{3z}$ forbids its symmetric part, thus $\sigma_{xy}=-\sigma_{yx}$, and the
transverse transport is described by a Hall conductivity $\sigma_{\mathrm{D}}=\left(
\sigma_{xy}-\sigma_{yx}\right)  /2$. The dipole Hall current reads
\begin{equation}
\boldsymbol{j}_{\mathrm{D}}=\sigma_{\mathrm{D}}\boldsymbol{E}_{\parallel}\times\boldsymbol{\hat{z}}, \label{dipolehall}
\end{equation}
where the dipole Hall conductivity is given by
\begin{equation}
\sigma_{\mathrm{D}}=\sum_{n}\int[d\boldsymbol{k}]f_{0}(\boldsymbol{k})\Upsilon_{\mathrm{D}}%
^{n}(\boldsymbol{k}),
\label{FHC}
\end{equation}
with $ \Upsilon_{\mathrm{D}}^{n}(\boldsymbol{k})=\epsilon_{ab}\Upsilon_{ab}^{n}(\boldsymbol{k})/2$ being the antisymmetric part of the dipole Berry curvature. Here $\epsilon$ is the Levi-Civita symbol.
As a time-reversal odd pseudoscalar ($zz$
component of a pseudotensor), $\sigma_{\mathrm{D}}$ is allowed by rotation and
primed improper rotations (which are combinations of time reversal with inversion, mirror or roto-reflection).

\emph{{\color{blue} Longitudinal orbital magnetoelectric response.}}-- The interlayer-dipole Hall current is also manifested as an in-plane orbital magnetization parallel to $E_{\parallel}$, as shown by the schematics in Figs.~\ref{fig:schematic}(c) and \ref{fig:schematic}(d). This is physically intuitive as a travelling charge dipole can induce an orbital magnetic moment \cite{Xiao2021CS}. To see this connection formally, one notices that the in-plane orbital magnetic moment operator
$
\boldsymbol{\hat{m}}=\frac{1}{4}(\boldsymbol{\hat{p}}\times\boldsymbol{\hat
{v}}-\boldsymbol{\hat{v}}\times\boldsymbol{\hat{p}})
$ can be recast into
$
\hat{\boldsymbol{m}} = \frac{d_{0}}{2} \hat{\boldsymbol{z}} \times \hat{\boldsymbol{j}},
%\label{eq:dhall_mag}
$
where $\boldsymbol{\hat{j}}$ is the aforementioned dipole current operator (details in \cite{supp}). 
This form of the in-plane orbital magnetic moment operator can also be obtained by rigorous quantum mechanical treatment of magnetic field effect in the continuum model of coupled twisted bilayers~\cite{Ashvin2019,Qin2021}, as shown in the Supplemental Information~\cite{supp}.

%figure-2: dipole Hall of filling -1 FM insulating state
\begin{figure}[ptb]
\includegraphics[width=0.48\textwidth]{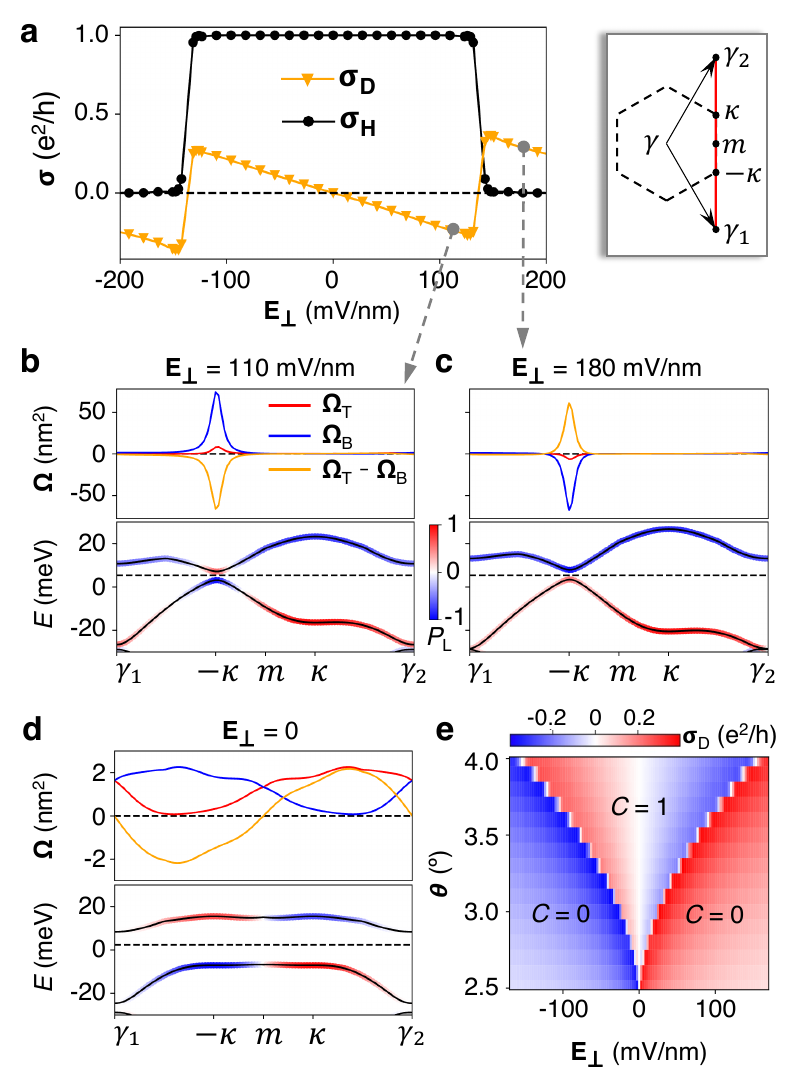}
\caption{Dipole Hall effect at $\nu=-1$ in tMoTe$_2$.
(a) Charge (black circles) and dipole (orange triangles) Hall conductivity as a function of $E_{\perp}$. Twist angle $\theta=3.9^{\circ}$.
(b-d) Berry curvatures (upper panel) and quasiparticle band dispersion (lower panel) along \textit{k} path $\gamma_1$ to $\gamma_2$ (c.f. inset for moir\'{e} Brillouin zone).
The band dispersion is color coded with the layer polarization: $P_l=1 $ $(-1)$ for Bloch states fully polarized on layer $T$ ($B$).
(e) Dipole Hall conductivity as a function of twist angle and $E_{\perp}$, where the abrupt sign changes mark the topological phase boundaries.
}%
\label{fig:filling1}%
\end{figure}

The equilibrium in-plane magnetization is prohibited by the $C_{3z}$ symmetry. Thus, the in-plane orbital magnetization appears from the
first order of electric field $M_{\parallel,a}=\chi_{ab}E_{\parallel,b}$.
For the examples to be discussed, $C_{3z}$ symmetry renders $\chi
_{xx}=\chi_{yy}$, whereas $M_xT$ symmetry forbids $\chi
_{xy},\chi_{yx}$, leaving the magnetoelectric response a longitudinal form,
\begin{equation}\label{M}
\boldsymbol{M}_{\parallel}=\frac{d_{0}}{2}\sigma_{\mathrm{D}}\boldsymbol{E}_{\parallel}.
\end{equation}
Namely, this longitudinal magnetoelectric response is equivalent to the intrinsic dipole Hall effect (details in \cite{supp}). 
Therefore, measuring the in-plane magnetization, by magneto-optical means or via the magnetic stray field, allows a contact-free detection of the dipole Hall effect. 
Moreover, an AC $E_{\parallel}$ field will drive an oscillating $\boldsymbol{M}_{\parallel}$ that adds on top of the equilibrium magnetization [Fig.~\ref{fig:schematic}(e)], such that the net magnetization can precess with a frequency upper bounded by the charge gap, reaching THz range for the examples below. 

Before proceeding to specific behaviors of the proposed effects, some comments are in order. 

First, the orbital magnetization in linear response to $E$ field in 2D insulators is a boundary-independent bulk thermodynamic quantity. Based on the in-plane orbital moment operator obtained by the quantum mechanical treatment and perturbation calculations on its response to in-plane electric field, we determine the in-plane intrinsic orbital magnetoelectric response unambiguously~\cite{supp}. Consistently, one can also obtain this response coefficient by using the Maxwell relation, which states that it also quantifies the linear response of in-plane electric polarization to in-plane magnetic field. We have performed direct perturbation calculation of the latter response and get the expected same result~\cite{supp}, further corroborating our theory for the in-plane orbital magnetoelectric response.
%First, the orbital magnetization in linear response to $E$ field in insulators is a boundary-independent bulk thermodynamic quantity. Consistently, one can obtain the in-plane intrinsic orbital magnetoelectric coefficient by using the Maxwell relation, which states that this coefficient also quantifies the linear response of in-plane electric polarization to in-plane magnetic field. Direct perturbation calculation of the in-plane polarization gives the expected same result, further confirming the correctness of our theory for the in-plane orbital magnetoelectric response.

Second, the measurable longitudinal orbital magnetoelectric susceptibility is found to be quantitatively equivalent to the bulk dipole Hall conductivity in terms of the conventional dipole current definition employed here, although there can be alternative definition of dipole current operator in the presence of interlayer coupling  (similar to the case of defining spin current in the presence of spin-orbit coupling \cite{Dimi2004}). On the one hand, it is reminded that our formulated magnetoelectric response is unambiguous, independent of the definition of dipole Hall current. On the other hand, the quantitative equivalence means that the in-plane magnetoelectric response renders a way to measure the bulk dipole Hall conductivity of the conventional dipole current.

Third, the interlayer tunneling plays a critical role in making relevant an in-plane orbital magnetoelectricity in the context of a double layer 2D system. It is the interplay of the interlayer tunneling and intralayer moiré potentials that underlies this magnetoelectric response, while setting either to zero will lead to a null response [c.f. Supplementary Fig.~S3].

%If an equilibrium out-of-plane magnetization is symmetry allowed, rotating the in-plane field may also enable magnetization precession [Fig.~\ref{fig:schematic}(e)], which has a novel orbital origin.

\emph{{\color{blue} Dipole Hall effect in the FM insulator at $\nu=-1$.}}-- We first demonstrate the dipole Hall conductivity at $\nu=-1$ for 3.9$^\circ$ tMoTe$_{2}$.
Figure~\ref{fig:filling1}(a) shows the variation of Hall conductivities with interlayer bias, from a Hartree-Fock mean-field calculation~\cite{supp}.
The black-circle (orange-triangle) curve represents the charge (dipole) Hall conductivity.
At small interlayer bias $E_{\perp}$, the FM is in a QAH state with Chern number $1$, while the dipole Hall conductivity vanishes at $E_{\perp}=0$ due to the $C_{2y}T$ symmetry.
A finite $E_{\perp}$ breaks the $C_{2y}T$ and switches on the dipole Hall effect. The FM transits from QAH to a topologically trivial state at a critical $E_{\perp}$, upon which the dipole Hall conductivity undergoes an abrupt change.

%figure-3: dipole Hall of filling-2 AFMz state
\begin{figure}[ptb]
\includegraphics[width=0.48\textwidth]{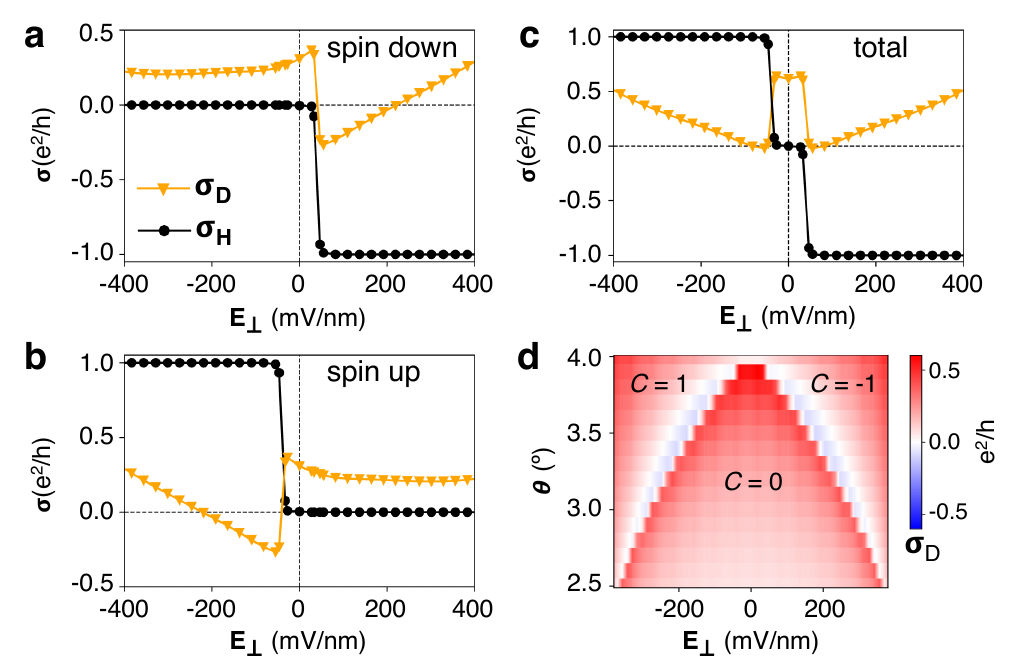}
\caption{Dipole Hall effect in the AFMz insulating state at $\nu=-2$ in tMoTe$_2$.
(a)-(c) Dipole Hall (black circles) and charge Hall conductivity (orange triangles) as functions of interlayer bias for (a) spin-down, (b) spin-up, and (c) total contributions, respectively.
The twist angle $\theta=3.9^\circ$
(d) Dipole Hall conductivity as a function of twist angle and interlayer bias.
}%
\label{fig:filling2}%
\end{figure}

The simultaneous abrupt change in the charge and dipole Hall conductivities is not a coincidence. The band geometric properties arising from layer pseudospin textures underly both quantities, which become noncontinuous at the band inversion induced by the interlayer bias. To see this, we introduce layer-projected Berry curvature $\Omega_l$ ($l=T, B$) which satisfies $\Omega_T+\Omega_B=\Omega$ and $\Omega_T-\Omega_B=\Upsilon_{\mathrm{D}}$, where $\Omega$ is the usual $k$-space Berry curvature. 
%Figure~\ref{fig:filling1}(d) displays the Berry curvatures and the Hartree-Fock band dispersion with interlayer-dipole texture at $E_{\perp}=0$. In this case the dipole Berry curvature exhibits an antisymmetric distribution about the origin in the Brillouin zone, and the dipole Hall effect is forbidden by $C_{2y}T$. When $E_{\perp}\neq 0$, 
Figures~\ref{fig:filling1}(b) and \ref{fig:filling1}(c) show the Berry curvatures (upper panel) and band dispersion (lower panel) before and after the topological phase transition, respectively. Notably, at these positive interlayer bias, the filled band predominantly occupies layer $B$, and $\Omega_B$ dominates over $\Omega_T$ in magnitude.
When $E_{\perp}$ crosses the transition point, band inversion occurs in the vicinity of $-\kappa$ point, where the Berry curvatures all get reversed.
As a result, an abrupt change of $\sigma_D$ accompanies the change of Chern number. The reversal of the associated in-plane orbital magnetization [Figs.~\ref{fig:schematic}(c) and \ref{fig:schematic}(d)] enables contact-free detection of this topological transition.
%In the topologically trivial region, as the bias increases, the interlayer hybridization weakens, and the dipole Hall conductivity gradually decreases.

The phase diagram for the dipole Hall conductivity as a function of twist angle and bias is presented as Fig.~\ref{fig:filling1}(e).
One observes that as the twist angle decreases, the critical bias for dipole-Hall jump decreases.
This is because as the twist angle decreases, the energy band also narrows, and the critical bias decreases accordingly.

\emph{{\color{blue} Dipole Hall effect in the AFM insulator at $\nu=-2$.}}-- Compared to the case of $\nu=-1$, the $\nu=-2$ AFMz state exhibits a different symmetry, where $C_{2y}T$ is replaced by $C_{2y}$ [Fig.~\ref{fig:schematic}(b)]. Therefore, a pronounced dipole Hall effect is present at $E_{\perp}=0$ [Fig.~\ref{fig:filling2}(c)].
%, which is well within the experimental capacity. 
%This is the first material example for the quantum solenoids \cite{Law2021} without Fermi surface, which is of orbital origin based on dipole Hall transport. 
% 0.21~$\mu_B/MUC$
As the interlayer bias is increased, the dipole Hall conductivity undergoes a sudden jump [Fig.~\ref{fig:filling2}(c)] upon the topological transition to the AFMz QAH state, reminiscent of the finding in the $\nu=-1$ FM configuration. We separately examine the contributions from the spin-down and spin-up channels to charge and dipole Hall conductivities, as shown in Figs.~\ref{fig:filling2}(a) and (b), respectively.
As the bias increases in the positive (negative) direction, the transition from trivial to QAH state occurs in the spin-down (spin-up) channel, whose contribution to the dipole Hall conductivity has a sudden sign change. The underlying picture is similar to the case of $\nu=-1$, where the band inversion in a spin channel leads to an abrupt change in both $\sigma_D$ and Chern
number (details in \cite{supp}).
%, \wy{the sign of dipole Hall conductivity has an abrupt change in the spin channel where topological  transition occurs.}
%which are explained schematically in Fig.~\ref{fig:filling2}(e) and (f).

The dependence of dipole Hall conductivity on twist angle and interlayer bias is shown in Fig.~\ref{fig:filling2}(d).
The topological phase diagram is  complementary to that of the $\nu=-1$ case, where critical $E_{\perp}$  decreases with twisting angle~\cite{fan2024Orbital}.
With contributions from both spin up and down carriers, $\sigma_D$ has larger magnitude here compared to the $\nu=-1$ case. Near zero interlayer bias, we find a modest $E_{\parallel} = 10^7$ V/m can generate a sizable in-plane orbital magnetization $M_{\parallel} \sim 0.01$ of $\mu_B/$nm$^2$.

\begin{figure}[ptb]
\includegraphics[width=0.48\textwidth]{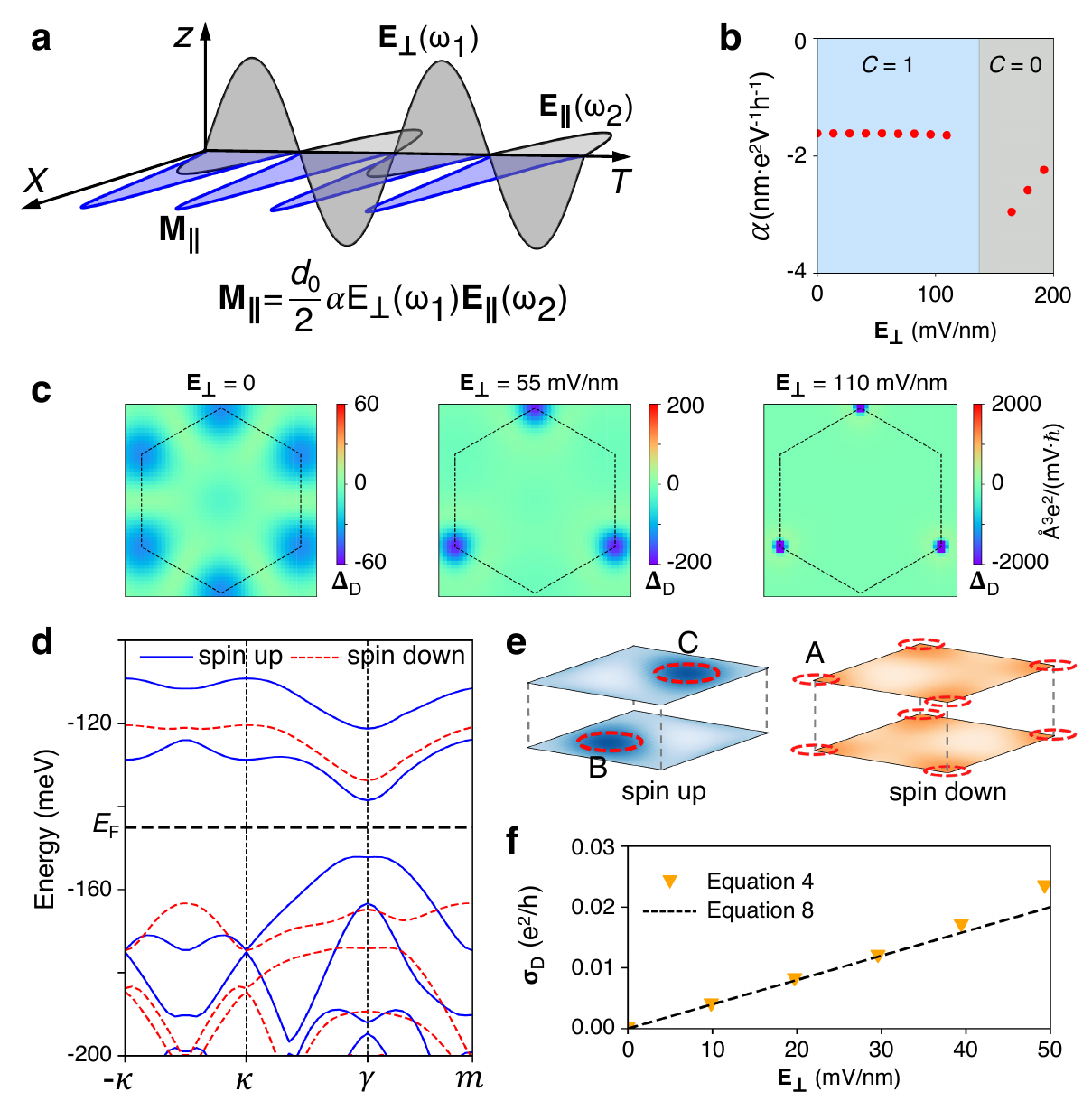}
\caption{Crossed nonlinear dipole Hall effect at $\nu=-1$ and $-3$ in tMoTe$_2$. (a) Schematic of the nonlinear orbital magnetoelectric response to two AC electric fields in-plane and out-of-plane. (b) The flux of dipole Berry curvature polarizability as function of $E_{\perp}$ at $\nu=-1$, in unit of $\frac{nm\cdot e^2}{V\cdot h}$. (c) Distinct distributions of dipole Berry curvature polarizability in the moir\'{e} Brillouin zone at three different $E_{\perp}$ in the QAH phase.
(d) Quasiparticle band dispersion of $\nu=-3$ ferrimagnetic state, for which the twist angle $\theta=3.5^\circ$. The blue-solid (red-dashed) lines represent spin-up (spin-down) bands. The black-dashed line denotes the Fermi level.
(e) Layer-resolved carrier distribution of this ferrimagnetic state in a moir\'e supercell. Two spin-up holes occupy the B and C moir\'{e} orbitals respectively, and one spin-down hole occupies the layer-hybridized A orbital.
(f) Dipole Hall conductivity as a function of interlayer bias. The black-dashed line is generated using $\sigma_{\mathrm{D}} = \alpha(0) E_{\perp}$.
}%
\label{fig:nonlinear}%
\end{figure}

\emph{\color{blue} Crossed nonlinear dipole Hall effect.}-- The study of nonlinear Hall effect is another recent focus of condensed matter physics \cite{Ma2019,Kang2019,Lu2021}. With $\sigma_D$ symmetry forbidden in the $\nu=-1$ state at $E_{\perp}=0$, its $E_{\perp}$ dependence in Fig.~\ref{fig:filling1}(a) implies a new type of nonlinearity --- intrinsic nonlinear dipole Hall effect [Fig.~\ref{fig:nonlinear}(a)].
We can define $\Delta_{\mathrm{D}}^n(\boldsymbol{k}) = \partial\Upsilon_{\mathrm{D}}^{n}(\boldsymbol{k})/\partial E_{\perp}$ as the {\it dipole Berry curvature polarizability} with respect to $E_{\perp}$, which is also a band geometric quantity: 
%$\delta \Upsilon_{\mathrm{D}}^{n}(\boldsymbol{k})=\Delta_{\mathrm{D}}^n(\boldsymbol{k}) E_{\perp}$, where~\cite{supp}
\begin{equation}
\begin{aligned}
\Delta_{\mathrm{D}}^n =& \frac{e \hbar}{2} \operatorname{Im} \sum_{m \neq n}
 \left[\frac{2 \boldsymbol{j}^{n m} \times \boldsymbol{v}^{m n}\left(p^n-p^m\right)}{\left(\varepsilon_n-\varepsilon_m\right)^3} \right.\\
 & \left. - \sum_{\ell \neq n} \frac{\left(\boldsymbol{j}^{\ell m} \times \boldsymbol{v}^{m n}+\boldsymbol{v}^{\ell m} \times \boldsymbol{j}^{m n}\right) p^{n \ell}}{\left(\varepsilon_n-\varepsilon_{\ell}\right)\left(\varepsilon_n-\varepsilon_m\right)^2} \right.\\
 & \left. + \sum_{\ell \neq m} \frac{\left(\boldsymbol{j}^{\ell n} \times \boldsymbol{v}^{n m}+\boldsymbol{v}^{\ell n} \times \boldsymbol{j}^{n m}\right) p^{m \ell}}{\left(\varepsilon_m-\varepsilon_{\ell}\right)\left(\varepsilon_n-\varepsilon_m\right)^2}\right] \label{polarizability}
\end{aligned}
\end{equation}
and its flux through filled bands,
\begin{equation}
\alpha (E_{\perp})= \sum_{n}\int[d \boldsymbol{k}] f_{0}(\boldsymbol{k}) \Delta_{\mathrm{D}}^n (\boldsymbol{k}).
\label{eq:dhall_alpha}
\end{equation}
Substituting them into Eq.~(\ref{dipolehall}),
we get a crossed nonlinear dipole current response,
\begin{equation}
\boldsymbol{j}_{\mathrm{D}} = \alpha(0) \boldsymbol{E}_{\|} \times \boldsymbol{E}_{\perp},
\label{eq:dhall_nonlinear}
\end{equation}
The concomitant nonlinear orbital magnetoelectric response is given by 
\begin{equation}
\boldsymbol{M}_{\parallel}=\frac{d_{0}}{2}\alpha(0)E_{\perp}\boldsymbol{E}_{\parallel}. \label{eq:EM_nonlinear}
\end{equation}
As shown in Fig.~\ref{fig:nonlinear}(a), when the two electric fields are AC, $M_{\parallel}$ have sum-frequency and difference-frequency components.

% figure-4: crossed nonlinear responses

Surprisingly, Figure~\ref{fig:filling1}(a) shows the simple scaling of the crossed nonlinear response given in Eq.~(\ref{eq:dhall_nonlinear}) and (\ref{eq:EM_nonlinear}) is protected over the entire QAH phase, until a sudden jump occurs as the signature of the topological transition. Over such broad range of $E_{\perp}$, the variation of the minibands [Fig.~2(b,d)] does lead to significant changes in the distribution of $\Delta_{\mathrm{D}}^n(\boldsymbol{k})$ in the moir\'e Brillouin zone, as shown in Fig.~4(c). Nevertheless, their flux over the filled Chern band remains a constant in the QAH phase, in stark contrast to that in the topological trivial phases [Fig.~4(b)]. This protected scaling throughout the QAH phase is observed at other twisting angles, and also for the AFMz configuration, while the actual flux can vary modestly with angle (details in \cite{supp}).

The nonlinear dipole Hall response is not limited to QAH state.
We showcase another example at filling factor $\nu=-3$, which hosts a topologically trivial ferrimagnetic insulating state, with two spin-up holes in the layer-polarized B and C orbitals and one spin-down hole in the layer-hybridized A orbital [Fig.~\ref{fig:nonlinear}(d) and \ref{fig:nonlinear}(e)]. Details of the Hartree-Fock calculation are given in~\cite{supp}.
Possessing $C_{2y}T$ symmetry as well [Fig.~\ref{fig:nonlinear}(e)], the linear dipole Hall effect is forbidden at zero interlayer bias, whereas the crossed nonlinear dipole Hall response is anticipated. This is confirmed by the calculated dipole Hall conductivity as a function of $E_{\perp}$ given in Fig.~\ref{fig:nonlinear}(f), which shows that the response is well captured by Eq.~(\ref{eq:dhall_nonlinear}) for $E_{\perp}$ range up to 40~mV/nm.

\emph{\color{blue} Discussion.}--%{\color{red}
We can also provide an alternative picture for the magnetoelectricity here. The bulk dipole Hall current tends to accumulate interlayer dipole on the boundary, which will get annihilated through interlayer tunneling. This tunneling current is inversely proportional to the dipole relaxation time and proportional to the dipole density. In steady state, the interlayer tunneling current on the edges will balance with the bulk dipole Hall current that corresponds to layer counter flows. Together they form a current loop that generates the in-plane magnetization in bulk.

Compared to the known magnetoelectric phenomena \cite{dong2015multiferroic,spaldin2019advances}, the magnetoelectricity from the intrinsic dipole Hall effect here has several features. The in-plane magnetoelectricity in 2D insulators here is solely of orbital origin, enabled by the spin-conserved interlayer tunneling across the twisted interface~\cite{Stauber2018,zhai2023Timereversal}. As an intrinsic effect in an insulator, the magnetoelectric response here is protected by a charge gap. This allows the phenomena to be explored in the AC regime, where the generated magnetization can adiabatically follow an AC electric field, generating magnetization oscillations at high frequencies upper bounded by the charge gap. This could potentially provide a mechanism to induce magnetic oscillations of ferromagnets at faster timescales. The form of crossed nonlinear magnetoelectric susceptibility further points to novel device functionalities, where the nonlinearity underlies logic operations, second harmonic generation, rectification, etc. 

In comparison with the magnon based functionalities also based on magnetic insulators, the intrinsic dipole Hall effect exhibits several complementary features. The magnon transport is typically driven by temperature gradient and magnetic field gradient~\cite{bauer2022magnonics}, while the dipole Hall current is a response directly to an in-plane electric field through band geometric effect. The response can also be dramatically tuned by the out-of-plane electric field, including the magnitude of the susceptibility and the abrupt sign change when tuning across the topological phase transition point. Exploring the transport of electric dipole and that of magnon in magnetic insulators represent distinct and complementary approaches to low-power-consumption information technologies, which could potentially be combined to exploit the advantages of both.
\emph{\color{blue} Acknowledgements.}--This work is supported by the Research Grant Council of Hong Kong SAR China (AoE/P-701/20, HKU SRFS2122-7S05, A-HKU705/21), the National Key R\&D Program of China (2020YFA0309600), and New Cornerstone Science Foundation. C.X. acknowledges support by the UM Start-up Grant (SRG2023-00033-IAPME).

% Data availability
%\emph{\color{blue} Data availability.}--The numerical data generated by the custom codes for this study that support the findings are available from the corresponding author on reasonable request.

% Code availability
%\emph{\color{blue} Code availability.}--The custom codes prepared for this study that support the findings are available from the corresponding author on reasonable request.

% cite references in bibliography file
\bibliographystyle{apsrev4-2} % comment to show article titles in references.
\bibliography{Dipole_Hall}

\begin{thebibliography}{45}%
\makeatletter
\providecommand \@ifxundefined [1]{%
 \@ifx{#1\undefined}
}%
\providecommand \@ifnum [1]{%
 \ifnum #1\expandafter \@firstoftwo
 \else \expandafter \@secondoftwo
 \fi
}%
\providecommand \@ifx [1]{%
 \ifx #1\expandafter \@firstoftwo
 \else \expandafter \@secondoftwo
 \fi
}%
\providecommand \natexlab [1]{#1}%
\providecommand \enquote  [1]{``#1''}%
\providecommand \bibnamefont  [1]{#1}%
\providecommand \bibfnamefont [1]{#1}%
\providecommand \citenamefont [1]{#1}%
\providecommand \href@noop [0]{\@secondoftwo}%
\providecommand \href [0]{\begingroup \@sanitize@url \@href}%
\providecommand \@href[1]{\@@startlink{#1}\@@href}%
\providecommand \@@href[1]{\endgroup#1\@@endlink}%
\providecommand \@sanitize@url [0]{\catcode `\\12\catcode `\$12\catcode
  `\&12\catcode `\#12\catcode `\^12\catcode `\_12\catcode `\%12\relax}%
\providecommand \@@startlink[1]{}%
\providecommand \@@endlink[0]{}%
\providecommand \url  [0]{\begingroup\@sanitize@url \@url }%
\providecommand \@url [1]{\endgroup\@href {#1}{\urlprefix }}%
\providecommand \urlprefix  [0]{URL }%
\providecommand \Eprint [0]{\href }%
\providecommand \doibase [0]{https://doi.org/}%
\providecommand \selectlanguage [0]{\@gobble}%
\providecommand \bibinfo  [0]{\@secondoftwo}%
\providecommand \bibfield  [0]{\@secondoftwo}%
\providecommand \translation [1]{[#1]}%
\providecommand \BibitemOpen [0]{}%
\providecommand \bibitemStop [0]{}%
\providecommand \bibitemNoStop [0]{.\EOS\space}%
\providecommand \EOS [0]{\spacefactor3000\relax}%
\providecommand \BibitemShut  [1]{\csname bibitem#1\endcsname}%
\let\auto@bib@innerbib\@empty
%</preamble>
\bibitem [{\citenamefont {Chumak}\ \emph {et~al.}(2015)\citenamefont {Chumak},
  \citenamefont {Vasyuchka}, \citenamefont {Serga},\ and\ \citenamefont
  {Hillebrands}}]{chumak2015Magnon}%
  \BibitemOpen
  \bibfield  {author} {\bibinfo {author} {\bibfnamefont {A.~V.}\ \bibnamefont
  {Chumak}}, \bibinfo {author} {\bibfnamefont {V.~I.}\ \bibnamefont
  {Vasyuchka}}, \bibinfo {author} {\bibfnamefont {A.~A.}\ \bibnamefont
  {Serga}},\ and\ \bibinfo {author} {\bibfnamefont {B.}~\bibnamefont
  {Hillebrands}},\ }\href {https://www.nature.com/articles/nphys3347}
  {\bibfield  {journal} {\bibinfo  {journal} {Nat. Phys.}\ }\textbf {\bibinfo
  {volume} {11}},\ \bibinfo {pages} {453} (\bibinfo {year} {2015})}\BibitemShut
  {NoStop}%
\bibitem [{\citenamefont {Yu}\ \emph {et~al.}(2018)\citenamefont {Yu},
  \citenamefont {Xiao},\ and\ \citenamefont {Pirro}}]{yu2018Magnon}%
  \BibitemOpen
  \bibfield  {author} {\bibinfo {author} {\bibfnamefont {H.}~\bibnamefont
  {Yu}}, \bibinfo {author} {\bibfnamefont {J.}~\bibnamefont {Xiao}},\ and\
  \bibinfo {author} {\bibfnamefont {P.}~\bibnamefont {Pirro}},\ }\href
  {https://www.sciencedirect.com/science/article/pii/S0304885317338118}
  {\bibfield  {journal} {\bibinfo  {journal} {J. Magn. Magn.}\ }\bibinfo
  {series} {Perspectives on Magnon Spintronics},\ \textbf {\bibinfo {volume}
  {450}},\ \bibinfo {pages} {1} (\bibinfo {year} {2018})}\BibitemShut {NoStop}%
\bibitem [{\citenamefont {Bauer}\ \emph {et~al.}(2023)\citenamefont {Bauer},
  \citenamefont {Tang}, \citenamefont {Iguchi}, \citenamefont {Xiao},
  \citenamefont {Shen}, \citenamefont {Zhong}, \citenamefont {Yu},
  \citenamefont {Rezende}, \citenamefont {Heremans},\ and\ \citenamefont
  {Uchida}}]{bauer2023Polarization}%
  \BibitemOpen
  \bibfield  {author} {\bibinfo {author} {\bibfnamefont {G.}~\bibnamefont
  {Bauer}}, \bibinfo {author} {\bibfnamefont {P.}~\bibnamefont {Tang}},
  \bibinfo {author} {\bibfnamefont {R.}~\bibnamefont {Iguchi}}, \bibinfo
  {author} {\bibfnamefont {J.}~\bibnamefont {Xiao}}, \bibinfo {author}
  {\bibfnamefont {K.}~\bibnamefont {Shen}}, \bibinfo {author} {\bibfnamefont
  {Z.}~\bibnamefont {Zhong}}, \bibinfo {author} {\bibfnamefont
  {T.}~\bibnamefont {Yu}}, \bibinfo {author} {\bibfnamefont {S.}~\bibnamefont
  {Rezende}}, \bibinfo {author} {\bibfnamefont {J.}~\bibnamefont {Heremans}},\
  and\ \bibinfo {author} {\bibfnamefont {K.}~\bibnamefont {Uchida}},\ }\href
  {https://link.aps.org/doi/10.1103/PhysRevApplied.20.050501} {\bibfield
  {journal} {\bibinfo  {journal} {Phys. Rev. Appl.}\ }\textbf {\bibinfo
  {volume} {20}},\ \bibinfo {pages} {050501} (\bibinfo {year}
  {2023})}\BibitemShut {NoStop}%
\bibitem [{\citenamefont {Shen}(2022)}]{shen2022Magnonferron}%
  \BibitemOpen
  \bibfield  {author} {\bibinfo {author} {\bibfnamefont {K.}~\bibnamefont
  {Shen}},\ }\href {https://link.aps.org/doi/10.1103/PhysRevB.106.104411}
  {\bibfield  {journal} {\bibinfo  {journal} {Phys. Rev. B}\ }\textbf {\bibinfo
  {volume} {106}},\ \bibinfo {pages} {104411} (\bibinfo {year}
  {2022})}\BibitemShut {NoStop}%
\bibitem [{\citenamefont {Chotorlishvili}\ \emph {et~al.}(2013)\citenamefont
  {Chotorlishvili}, \citenamefont {Khomeriki}, \citenamefont {Sukhov},
  \citenamefont {Ruffo},\ and\ \citenamefont
  {Berakdar}}]{chotorlishvili2013Dynamics}%
  \BibitemOpen
  \bibfield  {author} {\bibinfo {author} {\bibfnamefont {L.}~\bibnamefont
  {Chotorlishvili}}, \bibinfo {author} {\bibfnamefont {R.}~\bibnamefont
  {Khomeriki}}, \bibinfo {author} {\bibfnamefont {A.}~\bibnamefont {Sukhov}},
  \bibinfo {author} {\bibfnamefont {S.}~\bibnamefont {Ruffo}},\ and\ \bibinfo
  {author} {\bibfnamefont {J.}~\bibnamefont {Berakdar}},\ }\href
  {https://link.aps.org/doi/10.1103/PhysRevLett.111.117202} {\bibfield
  {journal} {\bibinfo  {journal} {Phys. Rev. Lett.}\ }\textbf {\bibinfo
  {volume} {111}},\ \bibinfo {pages} {117202} (\bibinfo {year}
  {2013})}\BibitemShut {NoStop}%
\bibitem [{\citenamefont {Bauer}\ \emph {et~al.}(2021)\citenamefont {Bauer},
  \citenamefont {Iguchi},\ and\ \citenamefont {Uchida}}]{bauer2021Theory}%
  \BibitemOpen
  \bibfield  {author} {\bibinfo {author} {\bibfnamefont {G.~E.~W.}\
  \bibnamefont {Bauer}}, \bibinfo {author} {\bibfnamefont {R.}~\bibnamefont
  {Iguchi}},\ and\ \bibinfo {author} {\bibfnamefont {K.-i.}\ \bibnamefont
  {Uchida}},\ }\href {https://link.aps.org/doi/10.1103/PhysRevLett.126.187603}
  {\bibfield  {journal} {\bibinfo  {journal} {Phys. Rev. Lett.}\ }\textbf
  {\bibinfo {volume} {126}},\ \bibinfo {pages} {187603} (\bibinfo {year}
  {2021})}\BibitemShut {NoStop}%
\bibitem [{\citenamefont {Chen}\ \emph {et~al.}(2021)\citenamefont {Chen},
  \citenamefont {Lan}, \citenamefont {Min},\ and\ \citenamefont
  {Xiao}}]{chen2021Narrow}%
  \BibitemOpen
  \bibfield  {author} {\bibinfo {author} {\bibfnamefont {G.}~\bibnamefont
  {Chen}}, \bibinfo {author} {\bibfnamefont {J.}~\bibnamefont {Lan}}, \bibinfo
  {author} {\bibfnamefont {T.}~\bibnamefont {Min}},\ and\ \bibinfo {author}
  {\bibfnamefont {J.}~\bibnamefont {Xiao}},\ }\href
  {https://iopscience.iop.org/article/10.1088/0256-307X/38/8/087701} {\bibfield
   {journal} {\bibinfo  {journal} {Chin. Phys. Lett.}\ }\textbf {\bibinfo
  {volume} {38}},\ \bibinfo {pages} {087701} (\bibinfo {year}
  {2021})}\BibitemShut {NoStop}%
\bibitem [{\citenamefont {Li}\ \emph {et~al.}(2021)\citenamefont {Li},
  \citenamefont {Stoica}, \citenamefont {Pa{\'s}ciak}, \citenamefont {Zhu},
  \citenamefont {Yuan}, \citenamefont {Yang}, \citenamefont {McCarter},
  \citenamefont {Das}, \citenamefont {Yadav}, \citenamefont {Park},
  \citenamefont {Dai}, \citenamefont {Lee}, \citenamefont {Ahn}, \citenamefont
  {Marks}, \citenamefont {Yu}, \citenamefont {Kadlec}, \citenamefont {Sato},
  \citenamefont {Hoffmann}, \citenamefont {Chollet}, \citenamefont {Kozina},
  \citenamefont {Nelson}, \citenamefont {Zhu}, \citenamefont {Walko},
  \citenamefont {Lindenberg}, \citenamefont {Evans}, \citenamefont {Chen},
  \citenamefont {Ramesh}, \citenamefont {Martin}, \citenamefont {Gopalan},
  \citenamefont {Freeland}, \citenamefont {Hlinka},\ and\ \citenamefont
  {Wen}}]{li2021Subterahertz}%
  \BibitemOpen
  \bibfield  {author} {\bibinfo {author} {\bibfnamefont {Q.}~\bibnamefont
  {Li}}, \bibinfo {author} {\bibfnamefont {V.~A.}\ \bibnamefont {Stoica}},
  \bibinfo {author} {\bibfnamefont {M.}~\bibnamefont {Pa{\'s}ciak}}, \bibinfo
  {author} {\bibfnamefont {Y.}~\bibnamefont {Zhu}}, \bibinfo {author}
  {\bibfnamefont {Y.}~\bibnamefont {Yuan}}, \bibinfo {author} {\bibfnamefont
  {T.}~\bibnamefont {Yang}}, \bibinfo {author} {\bibfnamefont {M.~R.}\
  \bibnamefont {McCarter}}, \bibinfo {author} {\bibfnamefont {S.}~\bibnamefont
  {Das}}, \bibinfo {author} {\bibfnamefont {A.~K.}\ \bibnamefont {Yadav}},
  \bibinfo {author} {\bibfnamefont {S.}~\bibnamefont {Park}}, \bibinfo {author}
  {\bibfnamefont {C.}~\bibnamefont {Dai}}, \bibinfo {author} {\bibfnamefont
  {H.~J.}\ \bibnamefont {Lee}}, \bibinfo {author} {\bibfnamefont
  {Y.}~\bibnamefont {Ahn}}, \bibinfo {author} {\bibfnamefont {S.~D.}\
  \bibnamefont {Marks}}, \bibinfo {author} {\bibfnamefont {S.}~\bibnamefont
  {Yu}}, \bibinfo {author} {\bibfnamefont {C.}~\bibnamefont {Kadlec}}, \bibinfo
  {author} {\bibfnamefont {T.}~\bibnamefont {Sato}}, \bibinfo {author}
  {\bibfnamefont {M.~C.}\ \bibnamefont {Hoffmann}}, \bibinfo {author}
  {\bibfnamefont {M.}~\bibnamefont {Chollet}}, \bibinfo {author} {\bibfnamefont
  {M.~E.}\ \bibnamefont {Kozina}}, \bibinfo {author} {\bibfnamefont
  {S.}~\bibnamefont {Nelson}}, \bibinfo {author} {\bibfnamefont
  {D.}~\bibnamefont {Zhu}}, \bibinfo {author} {\bibfnamefont {D.~A.}\
  \bibnamefont {Walko}}, \bibinfo {author} {\bibfnamefont {A.~M.}\ \bibnamefont
  {Lindenberg}}, \bibinfo {author} {\bibfnamefont {P.~G.}\ \bibnamefont
  {Evans}}, \bibinfo {author} {\bibfnamefont {L.-Q.}\ \bibnamefont {Chen}},
  \bibinfo {author} {\bibfnamefont {R.}~\bibnamefont {Ramesh}}, \bibinfo
  {author} {\bibfnamefont {L.~W.}\ \bibnamefont {Martin}}, \bibinfo {author}
  {\bibfnamefont {V.}~\bibnamefont {Gopalan}}, \bibinfo {author} {\bibfnamefont
  {J.~W.}\ \bibnamefont {Freeland}}, \bibinfo {author} {\bibfnamefont
  {J.}~\bibnamefont {Hlinka}},\ and\ \bibinfo {author} {\bibfnamefont
  {H.}~\bibnamefont {Wen}},\ }\href
  {https://www.nature.com/articles/s41586-021-03342-4} {\bibfield  {journal}
  {\bibinfo  {journal} {Nature}\ }\textbf {\bibinfo {volume} {592}},\ \bibinfo
  {pages} {376} (\bibinfo {year} {2021})}\BibitemShut {NoStop}%
\bibitem [{\citenamefont {Tang}\ \emph {et~al.}(2022)\citenamefont {Tang},
  \citenamefont {Iguchi}, \citenamefont {Uchida},\ and\ \citenamefont
  {Bauer}}]{tang2022Thermoelectric}%
  \BibitemOpen
  \bibfield  {author} {\bibinfo {author} {\bibfnamefont {P.}~\bibnamefont
  {Tang}}, \bibinfo {author} {\bibfnamefont {R.}~\bibnamefont {Iguchi}},
  \bibinfo {author} {\bibfnamefont {K.-i.}\ \bibnamefont {Uchida}},\ and\
  \bibinfo {author} {\bibfnamefont {G.~E.~W.}\ \bibnamefont {Bauer}},\ }\href
  {https://link.aps.org/doi/10.1103/PhysRevLett.128.047601} {\bibfield
  {journal} {\bibinfo  {journal} {Phys. Rev. Lett.}\ }\textbf {\bibinfo
  {volume} {128}},\ \bibinfo {pages} {047601} (\bibinfo {year}
  {2022})}\BibitemShut {NoStop}%
\bibitem [{\citenamefont {Weston}\ \emph {et~al.}(2022)\citenamefont {Weston},
  \citenamefont {Castanon}, \citenamefont {Enaldiev}, \citenamefont {Ferreira},
  \citenamefont {Bhattacharjee}, \citenamefont {Xu}, \citenamefont
  {{Corte-Le{\'o}n}}, \citenamefont {Wu}, \citenamefont {Clark}, \citenamefont
  {Summerfield}, \citenamefont {Hashimoto}, \citenamefont {Gao}, \citenamefont
  {Wang}, \citenamefont {Hamer}, \citenamefont {Read}, \citenamefont
  {Fumagalli}, \citenamefont {Kretinin}, \citenamefont {Haigh}, \citenamefont
  {Kazakova}, \citenamefont {Geim}, \citenamefont {Fal'ko},\ and\ \citenamefont
  {Gorbachev}}]{weston2022Interfacial}%
  \BibitemOpen
  \bibfield  {author} {\bibinfo {author} {\bibfnamefont {A.}~\bibnamefont
  {Weston}}, \bibinfo {author} {\bibfnamefont {E.~G.}\ \bibnamefont
  {Castanon}}, \bibinfo {author} {\bibfnamefont {V.}~\bibnamefont {Enaldiev}},
  \bibinfo {author} {\bibfnamefont {F.}~\bibnamefont {Ferreira}}, \bibinfo
  {author} {\bibfnamefont {S.}~\bibnamefont {Bhattacharjee}}, \bibinfo {author}
  {\bibfnamefont {S.}~\bibnamefont {Xu}}, \bibinfo {author} {\bibfnamefont
  {H.}~\bibnamefont {{Corte-Le{\'o}n}}}, \bibinfo {author} {\bibfnamefont
  {Z.}~\bibnamefont {Wu}}, \bibinfo {author} {\bibfnamefont {N.}~\bibnamefont
  {Clark}}, \bibinfo {author} {\bibfnamefont {A.}~\bibnamefont {Summerfield}},
  \bibinfo {author} {\bibfnamefont {T.}~\bibnamefont {Hashimoto}}, \bibinfo
  {author} {\bibfnamefont {Y.}~\bibnamefont {Gao}}, \bibinfo {author}
  {\bibfnamefont {W.}~\bibnamefont {Wang}}, \bibinfo {author} {\bibfnamefont
  {M.}~\bibnamefont {Hamer}}, \bibinfo {author} {\bibfnamefont
  {H.}~\bibnamefont {Read}}, \bibinfo {author} {\bibfnamefont {L.}~\bibnamefont
  {Fumagalli}}, \bibinfo {author} {\bibfnamefont {A.~V.}\ \bibnamefont
  {Kretinin}}, \bibinfo {author} {\bibfnamefont {S.~J.}\ \bibnamefont {Haigh}},
  \bibinfo {author} {\bibfnamefont {O.}~\bibnamefont {Kazakova}}, \bibinfo
  {author} {\bibfnamefont {A.~K.}\ \bibnamefont {Geim}}, \bibinfo {author}
  {\bibfnamefont {V.~I.}\ \bibnamefont {Fal'ko}},\ and\ \bibinfo {author}
  {\bibfnamefont {R.}~\bibnamefont {Gorbachev}},\ }\href
  {https://www.nature.com/articles/s41565-022-01072-w} {\bibfield  {journal}
  {\bibinfo  {journal} {Nat. Nanotechnol.}\ }\textbf {\bibinfo {volume} {17}},\
  \bibinfo {pages} {390} (\bibinfo {year} {2022})}\BibitemShut {NoStop}%
\bibitem [{\citenamefont {Wang}\ \emph {et~al.}(2022)\citenamefont {Wang},
  \citenamefont {Yasuda}, \citenamefont {Zhang}, \citenamefont {Liu},
  \citenamefont {Watanabe}, \citenamefont {Taniguchi}, \citenamefont {Hone},
  \citenamefont {Fu},\ and\ \citenamefont
  {{Jarillo-Herrero}}}]{wang2022Interfacial}%
  \BibitemOpen
  \bibfield  {author} {\bibinfo {author} {\bibfnamefont {X.}~\bibnamefont
  {Wang}}, \bibinfo {author} {\bibfnamefont {K.}~\bibnamefont {Yasuda}},
  \bibinfo {author} {\bibfnamefont {Y.}~\bibnamefont {Zhang}}, \bibinfo
  {author} {\bibfnamefont {S.}~\bibnamefont {Liu}}, \bibinfo {author}
  {\bibfnamefont {K.}~\bibnamefont {Watanabe}}, \bibinfo {author}
  {\bibfnamefont {T.}~\bibnamefont {Taniguchi}}, \bibinfo {author}
  {\bibfnamefont {J.}~\bibnamefont {Hone}}, \bibinfo {author} {\bibfnamefont
  {L.}~\bibnamefont {Fu}},\ and\ \bibinfo {author} {\bibfnamefont
  {P.}~\bibnamefont {{Jarillo-Herrero}}},\ }\href
  {https://www.nature.com/articles/s41565-021-01059-z} {\bibfield  {journal}
  {\bibinfo  {journal} {Nat. Nanotechnol.}\ }\textbf {\bibinfo {volume} {17}},\
  \bibinfo {pages} {367} (\bibinfo {year} {2022})}\BibitemShut {NoStop}%
\bibitem [{\citenamefont {Anderson}\ \emph {et~al.}(2023)\citenamefont
  {Anderson}, \citenamefont {Fan}, \citenamefont {Cai}, \citenamefont
  {Holtzmann}, \citenamefont {Taniguchi}, \citenamefont {Watanabe},
  \citenamefont {Xiao}, \citenamefont {Yao},\ and\ \citenamefont
  {Xu}}]{anderson2023Programming}%
  \BibitemOpen
  \bibfield  {author} {\bibinfo {author} {\bibfnamefont {E.}~\bibnamefont
  {Anderson}}, \bibinfo {author} {\bibfnamefont {F.-R.}\ \bibnamefont {Fan}},
  \bibinfo {author} {\bibfnamefont {J.}~\bibnamefont {Cai}}, \bibinfo {author}
  {\bibfnamefont {W.}~\bibnamefont {Holtzmann}}, \bibinfo {author}
  {\bibfnamefont {T.}~\bibnamefont {Taniguchi}}, \bibinfo {author}
  {\bibfnamefont {K.}~\bibnamefont {Watanabe}}, \bibinfo {author}
  {\bibfnamefont {D.}~\bibnamefont {Xiao}}, \bibinfo {author} {\bibfnamefont
  {W.}~\bibnamefont {Yao}},\ and\ \bibinfo {author} {\bibfnamefont
  {X.}~\bibnamefont {Xu}},\ }\href
  {https://www.science.org/doi/10.1126/science.adg4268} {\bibfield  {journal}
  {\bibinfo  {journal} {Science}\ }\textbf {\bibinfo {volume} {381}},\ \bibinfo
  {pages} {325} (\bibinfo {year} {2023})}\BibitemShut {NoStop}%
\bibitem [{\citenamefont {Cai}\ \emph {et~al.}(2023)\citenamefont {Cai},
  \citenamefont {Anderson}, \citenamefont {Wang}, \citenamefont {Zhang},
  \citenamefont {Liu}, \citenamefont {Holtzmann}, \citenamefont {Zhang},
  \citenamefont {Fan}, \citenamefont {Taniguchi}, \citenamefont {Watanabe},
  \citenamefont {Ran}, \citenamefont {Cao}, \citenamefont {Fu}, \citenamefont
  {Xiao}, \citenamefont {Yao},\ and\ \citenamefont {Xu}}]{cai2023Signatures}%
  \BibitemOpen
  \bibfield  {author} {\bibinfo {author} {\bibfnamefont {J.}~\bibnamefont
  {Cai}}, \bibinfo {author} {\bibfnamefont {E.}~\bibnamefont {Anderson}},
  \bibinfo {author} {\bibfnamefont {C.}~\bibnamefont {Wang}}, \bibinfo {author}
  {\bibfnamefont {X.}~\bibnamefont {Zhang}}, \bibinfo {author} {\bibfnamefont
  {X.}~\bibnamefont {Liu}}, \bibinfo {author} {\bibfnamefont {W.}~\bibnamefont
  {Holtzmann}}, \bibinfo {author} {\bibfnamefont {Y.}~\bibnamefont {Zhang}},
  \bibinfo {author} {\bibfnamefont {F.-R.}\ \bibnamefont {Fan}}, \bibinfo
  {author} {\bibfnamefont {T.}~\bibnamefont {Taniguchi}}, \bibinfo {author}
  {\bibfnamefont {K.}~\bibnamefont {Watanabe}}, \bibinfo {author}
  {\bibfnamefont {Y.}~\bibnamefont {Ran}}, \bibinfo {author} {\bibfnamefont
  {T.}~\bibnamefont {Cao}}, \bibinfo {author} {\bibfnamefont {L.}~\bibnamefont
  {Fu}}, \bibinfo {author} {\bibfnamefont {D.}~\bibnamefont {Xiao}}, \bibinfo
  {author} {\bibfnamefont {W.}~\bibnamefont {Yao}},\ and\ \bibinfo {author}
  {\bibfnamefont {X.}~\bibnamefont {Xu}},\ }\href
  {https://www.nature.com/articles/s41586-023-06289-w} {\bibfield  {journal}
  {\bibinfo  {journal} {Nature}\ }\textbf {\bibinfo {volume} {622}},\ \bibinfo
  {pages} {63} (\bibinfo {year} {2023})}\BibitemShut {NoStop}%
\bibitem [{\citenamefont {Park}\ \emph {et~al.}(2023)\citenamefont {Park},
  \citenamefont {Cai}, \citenamefont {Anderson}, \citenamefont {Zhang},
  \citenamefont {Zhu}, \citenamefont {Liu}, \citenamefont {Wang}, \citenamefont
  {Holtzmann}, \citenamefont {Hu}, \citenamefont {Liu}, \citenamefont
  {Taniguchi}, \citenamefont {Watanabe}, \citenamefont {Chu}, \citenamefont
  {Cao}, \citenamefont {Fu}, \citenamefont {Yao}, \citenamefont {Chang},
  \citenamefont {Cobden}, \citenamefont {Xiao},\ and\ \citenamefont
  {Xu}}]{park2023Observation}%
  \BibitemOpen
  \bibfield  {author} {\bibinfo {author} {\bibfnamefont {H.}~\bibnamefont
  {Park}}, \bibinfo {author} {\bibfnamefont {J.}~\bibnamefont {Cai}}, \bibinfo
  {author} {\bibfnamefont {E.}~\bibnamefont {Anderson}}, \bibinfo {author}
  {\bibfnamefont {Y.}~\bibnamefont {Zhang}}, \bibinfo {author} {\bibfnamefont
  {J.}~\bibnamefont {Zhu}}, \bibinfo {author} {\bibfnamefont {X.}~\bibnamefont
  {Liu}}, \bibinfo {author} {\bibfnamefont {C.}~\bibnamefont {Wang}}, \bibinfo
  {author} {\bibfnamefont {W.}~\bibnamefont {Holtzmann}}, \bibinfo {author}
  {\bibfnamefont {C.}~\bibnamefont {Hu}}, \bibinfo {author} {\bibfnamefont
  {Z.}~\bibnamefont {Liu}}, \bibinfo {author} {\bibfnamefont {T.}~\bibnamefont
  {Taniguchi}}, \bibinfo {author} {\bibfnamefont {K.}~\bibnamefont {Watanabe}},
  \bibinfo {author} {\bibfnamefont {J.-H.}\ \bibnamefont {Chu}}, \bibinfo
  {author} {\bibfnamefont {T.}~\bibnamefont {Cao}}, \bibinfo {author}
  {\bibfnamefont {L.}~\bibnamefont {Fu}}, \bibinfo {author} {\bibfnamefont
  {W.}~\bibnamefont {Yao}}, \bibinfo {author} {\bibfnamefont {C.-Z.}\
  \bibnamefont {Chang}}, \bibinfo {author} {\bibfnamefont {D.}~\bibnamefont
  {Cobden}}, \bibinfo {author} {\bibfnamefont {D.}~\bibnamefont {Xiao}},\ and\
  \bibinfo {author} {\bibfnamefont {X.}~\bibnamefont {Xu}},\ }\href
  {https://www.nature.com/articles/s41586-023-06536-0} {\bibfield  {journal}
  {\bibinfo  {journal} {Nature}\ }\textbf {\bibinfo {volume} {622}},\ \bibinfo
  {pages} {74} (\bibinfo {year} {2023})}\BibitemShut {NoStop}%
\bibitem [{\citenamefont {Xu}\ \emph {et~al.}(2023)\citenamefont {Xu},
  \citenamefont {Sun}, \citenamefont {Jia}, \citenamefont {Liu}, \citenamefont
  {Xu}, \citenamefont {Li}, \citenamefont {Gu}, \citenamefont {Watanabe},
  \citenamefont {Taniguchi}, \citenamefont {Tong}, \citenamefont {Jia},
  \citenamefont {Shi}, \citenamefont {Jiang}, \citenamefont {Zhang},
  \citenamefont {Liu},\ and\ \citenamefont {Li}}]{xu2023Observation}%
  \BibitemOpen
  \bibfield  {author} {\bibinfo {author} {\bibfnamefont {F.}~\bibnamefont
  {Xu}}, \bibinfo {author} {\bibfnamefont {Z.}~\bibnamefont {Sun}}, \bibinfo
  {author} {\bibfnamefont {T.}~\bibnamefont {Jia}}, \bibinfo {author}
  {\bibfnamefont {C.}~\bibnamefont {Liu}}, \bibinfo {author} {\bibfnamefont
  {C.}~\bibnamefont {Xu}}, \bibinfo {author} {\bibfnamefont {C.}~\bibnamefont
  {Li}}, \bibinfo {author} {\bibfnamefont {Y.}~\bibnamefont {Gu}}, \bibinfo
  {author} {\bibfnamefont {K.}~\bibnamefont {Watanabe}}, \bibinfo {author}
  {\bibfnamefont {T.}~\bibnamefont {Taniguchi}}, \bibinfo {author}
  {\bibfnamefont {B.}~\bibnamefont {Tong}}, \bibinfo {author} {\bibfnamefont
  {J.}~\bibnamefont {Jia}}, \bibinfo {author} {\bibfnamefont {Z.}~\bibnamefont
  {Shi}}, \bibinfo {author} {\bibfnamefont {S.}~\bibnamefont {Jiang}}, \bibinfo
  {author} {\bibfnamefont {Y.}~\bibnamefont {Zhang}}, \bibinfo {author}
  {\bibfnamefont {X.}~\bibnamefont {Liu}},\ and\ \bibinfo {author}
  {\bibfnamefont {T.}~\bibnamefont {Li}},\ }\href
  {https://link.aps.org/doi/10.1103/PhysRevX.13.031037} {\bibfield  {journal}
  {\bibinfo  {journal} {Phys. Rev. X}\ }\textbf {\bibinfo {volume} {13}},\
  \bibinfo {pages} {031037} (\bibinfo {year} {2023})}\BibitemShut {NoStop}%
\bibitem [{\citenamefont {Zeng}\ \emph {et~al.}(2023)\citenamefont {Zeng},
  \citenamefont {Xia}, \citenamefont {Kang}, \citenamefont {Zhu}, \citenamefont
  {Kn{\"u}ppel}, \citenamefont {Vaswani}, \citenamefont {Watanabe},
  \citenamefont {Taniguchi}, \citenamefont {Mak},\ and\ \citenamefont
  {Shan}}]{zeng2023Thermodynamica}%
  \BibitemOpen
  \bibfield  {author} {\bibinfo {author} {\bibfnamefont {Y.}~\bibnamefont
  {Zeng}}, \bibinfo {author} {\bibfnamefont {Z.}~\bibnamefont {Xia}}, \bibinfo
  {author} {\bibfnamefont {K.}~\bibnamefont {Kang}}, \bibinfo {author}
  {\bibfnamefont {J.}~\bibnamefont {Zhu}}, \bibinfo {author} {\bibfnamefont
  {P.}~\bibnamefont {Kn{\"u}ppel}}, \bibinfo {author} {\bibfnamefont
  {C.}~\bibnamefont {Vaswani}}, \bibinfo {author} {\bibfnamefont
  {K.}~\bibnamefont {Watanabe}}, \bibinfo {author} {\bibfnamefont
  {T.}~\bibnamefont {Taniguchi}}, \bibinfo {author} {\bibfnamefont {K.~F.}\
  \bibnamefont {Mak}},\ and\ \bibinfo {author} {\bibfnamefont {J.}~\bibnamefont
  {Shan}},\ }\href {https://www.nature.com/articles/s41586-023-06452-3}
  {\bibfield  {journal} {\bibinfo  {journal} {Nature}\ }\textbf {\bibinfo
  {volume} {622}},\ \bibinfo {pages} {69} (\bibinfo {year} {2023})}\BibitemShut
  {NoStop}%
\bibitem [{\citenamefont {Li}\ and\ \citenamefont {Wu}(2017)}]{li2017Binary}%
  \BibitemOpen
  \bibfield  {author} {\bibinfo {author} {\bibfnamefont {L.}~\bibnamefont
  {Li}}\ and\ \bibinfo {author} {\bibfnamefont {M.}~\bibnamefont {Wu}},\ }\href
  {https://doi.org/10.1021/acsnano.7b02756} {\bibfield  {journal} {\bibinfo
  {journal} {ACS Nano}\ }\textbf {\bibinfo {volume} {11}},\ \bibinfo {pages}
  {6382} (\bibinfo {year} {2017})}\BibitemShut {NoStop}%
\bibitem [{\citenamefont {Zhao}\ \emph {et~al.}(2021)\citenamefont {Zhao},
  \citenamefont {Xiao},\ and\ \citenamefont {Yao}}]{zhao2021Universal}%
  \BibitemOpen
  \bibfield  {author} {\bibinfo {author} {\bibfnamefont {P.}~\bibnamefont
  {Zhao}}, \bibinfo {author} {\bibfnamefont {C.}~\bibnamefont {Xiao}},\ and\
  \bibinfo {author} {\bibfnamefont {W.}~\bibnamefont {Yao}},\ }\href
  {https://www.nature.com/articles/s41699-021-00221-4} {\bibfield  {journal}
  {\bibinfo  {journal} {npj 2D Mater. Appl.}\ }\textbf {\bibinfo {volume}
  {5}},\ \bibinfo {pages} {1} (\bibinfo {year} {2021})}\BibitemShut {NoStop}%
\bibitem [{\citenamefont {Liang}\ \emph {et~al.}(2022)\citenamefont {Liang},
  \citenamefont {Yang}, \citenamefont {Wu}, \citenamefont {Dadap},
  \citenamefont {Watanabe}, \citenamefont {Taniguchi},\ and\ \citenamefont
  {Ye}}]{liang2022Optically}%
  \BibitemOpen
  \bibfield  {author} {\bibinfo {author} {\bibfnamefont {J.}~\bibnamefont
  {Liang}}, \bibinfo {author} {\bibfnamefont {D.}~\bibnamefont {Yang}},
  \bibinfo {author} {\bibfnamefont {J.}~\bibnamefont {Wu}}, \bibinfo {author}
  {\bibfnamefont {J.~I.}\ \bibnamefont {Dadap}}, \bibinfo {author}
  {\bibfnamefont {K.}~\bibnamefont {Watanabe}}, \bibinfo {author}
  {\bibfnamefont {T.}~\bibnamefont {Taniguchi}},\ and\ \bibinfo {author}
  {\bibfnamefont {Z.}~\bibnamefont {Ye}},\ }\href
  {https://link.aps.org/doi/10.1103/PhysRevX.12.041005} {\bibfield  {journal}
  {\bibinfo  {journal} {Phys. Rev. X}\ }\textbf {\bibinfo {volume} {12}},\
  \bibinfo {pages} {041005} (\bibinfo {year} {2022})}\BibitemShut {NoStop}%
\bibitem [{\citenamefont {Rog{\'e}e}\ \emph {et~al.}(2022)\citenamefont
  {Rog{\'e}e}, \citenamefont {Wang}, \citenamefont {Zhang}, \citenamefont
  {Cai}, \citenamefont {Wang}, \citenamefont {Chhowalla}, \citenamefont {Ji},\
  and\ \citenamefont {Lau}}]{rogee2022Ferroelectricity}%
  \BibitemOpen
  \bibfield  {author} {\bibinfo {author} {\bibfnamefont {L.}~\bibnamefont
  {Rog{\'e}e}}, \bibinfo {author} {\bibfnamefont {L.}~\bibnamefont {Wang}},
  \bibinfo {author} {\bibfnamefont {Y.}~\bibnamefont {Zhang}}, \bibinfo
  {author} {\bibfnamefont {S.}~\bibnamefont {Cai}}, \bibinfo {author}
  {\bibfnamefont {P.}~\bibnamefont {Wang}}, \bibinfo {author} {\bibfnamefont
  {M.}~\bibnamefont {Chhowalla}}, \bibinfo {author} {\bibfnamefont
  {W.}~\bibnamefont {Ji}},\ and\ \bibinfo {author} {\bibfnamefont {S.~P.}\
  \bibnamefont {Lau}},\ }\href
  {https://www.science.org/doi/10.1126/science.abm5734} {\bibfield  {journal}
  {\bibinfo  {journal} {Science}\ }\textbf {\bibinfo {volume} {376}},\ \bibinfo
  {pages} {973} (\bibinfo {year} {2022})}\BibitemShut {NoStop}%
\bibitem [{\citenamefont {Vizner~Stern}\ \emph {et~al.}(2021)\citenamefont
  {Vizner~Stern}, \citenamefont {Waschitz}, \citenamefont {Cao}, \citenamefont
  {Nevo}, \citenamefont {Watanabe}, \citenamefont {Taniguchi}, \citenamefont
  {Sela}, \citenamefont {Urbakh}, \citenamefont {Hod},\ and\ \citenamefont
  {Ben~Shalom}}]{viznerstern2021Interfacial}%
  \BibitemOpen
  \bibfield  {author} {\bibinfo {author} {\bibfnamefont {M.}~\bibnamefont
  {Vizner~Stern}}, \bibinfo {author} {\bibfnamefont {Y.}~\bibnamefont
  {Waschitz}}, \bibinfo {author} {\bibfnamefont {W.}~\bibnamefont {Cao}},
  \bibinfo {author} {\bibfnamefont {I.}~\bibnamefont {Nevo}}, \bibinfo {author}
  {\bibfnamefont {K.}~\bibnamefont {Watanabe}}, \bibinfo {author}
  {\bibfnamefont {T.}~\bibnamefont {Taniguchi}}, \bibinfo {author}
  {\bibfnamefont {E.}~\bibnamefont {Sela}}, \bibinfo {author} {\bibfnamefont
  {M.}~\bibnamefont {Urbakh}}, \bibinfo {author} {\bibfnamefont
  {O.}~\bibnamefont {Hod}},\ and\ \bibinfo {author} {\bibfnamefont
  {M.}~\bibnamefont {Ben~Shalom}},\ }\href
  {https://www.science.org/doi/10.1126/science.abe8177} {\bibfield  {journal}
  {\bibinfo  {journal} {Science}\ }\textbf {\bibinfo {volume} {372}},\ \bibinfo
  {pages} {1462} (\bibinfo {year} {2021})}\BibitemShut {NoStop}%
\bibitem [{\citenamefont {Woods}\ \emph {et~al.}(2021)\citenamefont {Woods},
  \citenamefont {Ares}, \citenamefont {{Nevison-Andrews}}, \citenamefont
  {Holwill}, \citenamefont {Fabregas}, \citenamefont {Guinea}, \citenamefont
  {Geim}, \citenamefont {Novoselov}, \citenamefont {Walet},\ and\ \citenamefont
  {Fumagalli}}]{woods2021Chargepolarized}%
  \BibitemOpen
  \bibfield  {author} {\bibinfo {author} {\bibfnamefont {C.~R.}\ \bibnamefont
  {Woods}}, \bibinfo {author} {\bibfnamefont {P.}~\bibnamefont {Ares}},
  \bibinfo {author} {\bibfnamefont {H.}~\bibnamefont {{Nevison-Andrews}}},
  \bibinfo {author} {\bibfnamefont {M.~J.}\ \bibnamefont {Holwill}}, \bibinfo
  {author} {\bibfnamefont {R.}~\bibnamefont {Fabregas}}, \bibinfo {author}
  {\bibfnamefont {F.}~\bibnamefont {Guinea}}, \bibinfo {author} {\bibfnamefont
  {A.~K.}\ \bibnamefont {Geim}}, \bibinfo {author} {\bibfnamefont {K.~S.}\
  \bibnamefont {Novoselov}}, \bibinfo {author} {\bibfnamefont {N.~R.}\
  \bibnamefont {Walet}},\ and\ \bibinfo {author} {\bibfnamefont
  {L.}~\bibnamefont {Fumagalli}},\ }\href
  {https://www.nature.com/articles/s41467-020-20667-2} {\bibfield  {journal}
  {\bibinfo  {journal} {Nat. Commun.}\ }\textbf {\bibinfo {volume} {12}},\
  \bibinfo {pages} {347} (\bibinfo {year} {2021})}\BibitemShut {NoStop}%
\bibitem [{\citenamefont {Yasuda}\ \emph {et~al.}(2021)\citenamefont {Yasuda},
  \citenamefont {Wang}, \citenamefont {Watanabe}, \citenamefont {Taniguchi},\
  and\ \citenamefont {{Jarillo-Herrero}}}]{yasuda2021Stackingengineered}%
  \BibitemOpen
  \bibfield  {author} {\bibinfo {author} {\bibfnamefont {K.}~\bibnamefont
  {Yasuda}}, \bibinfo {author} {\bibfnamefont {X.}~\bibnamefont {Wang}},
  \bibinfo {author} {\bibfnamefont {K.}~\bibnamefont {Watanabe}}, \bibinfo
  {author} {\bibfnamefont {T.}~\bibnamefont {Taniguchi}},\ and\ \bibinfo
  {author} {\bibfnamefont {P.}~\bibnamefont {{Jarillo-Herrero}}},\ }\href
  {https://www.science.org/doi/10.1126/science.abd3230} {\bibfield  {journal}
  {\bibinfo  {journal} {Science}\ }\textbf {\bibinfo {volume} {372}},\ \bibinfo
  {pages} {1458} (\bibinfo {year} {2021})}\BibitemShut {NoStop}%
\bibitem [{\citenamefont {Wu}\ \emph {et~al.}(2019)\citenamefont {Wu},
  \citenamefont {Lovorn}, \citenamefont {Tutuc}, \citenamefont {Martin},\ and\
  \citenamefont {MacDonald}}]{wu2019Topological}%
  \BibitemOpen
  \bibfield  {author} {\bibinfo {author} {\bibfnamefont {F.}~\bibnamefont
  {Wu}}, \bibinfo {author} {\bibfnamefont {T.}~\bibnamefont {Lovorn}}, \bibinfo
  {author} {\bibfnamefont {E.}~\bibnamefont {Tutuc}}, \bibinfo {author}
  {\bibfnamefont {I.}~\bibnamefont {Martin}},\ and\ \bibinfo {author}
  {\bibfnamefont {A.~H.}\ \bibnamefont {MacDonald}},\ }\href
  {https://link.aps.org/doi/10.1103/PhysRevLett.122.086402} {\bibfield
  {journal} {\bibinfo  {journal} {Phys. Rev. Lett.}\ }\textbf {\bibinfo
  {volume} {122}} (\bibinfo {year} {2019})}\BibitemShut {NoStop}%
\bibitem [{\citenamefont {Yu}\ \emph {et~al.}(2020)\citenamefont {Yu},
  \citenamefont {Chen},\ and\ \citenamefont {Yao}}]{yu2020Giant}%
  \BibitemOpen
  \bibfield  {author} {\bibinfo {author} {\bibfnamefont {H.}~\bibnamefont
  {Yu}}, \bibinfo {author} {\bibfnamefont {M.}~\bibnamefont {Chen}},\ and\
  \bibinfo {author} {\bibfnamefont {W.}~\bibnamefont {Yao}},\ }\href
  {https://academic.oup.com/nsr/article/7/1/12/5549053} {\bibfield  {journal}
  {\bibinfo  {journal} {Natl. Sci. Rev.}\ }\textbf {\bibinfo {volume} {7}},\
  \bibinfo {pages} {12} (\bibinfo {year} {2020})}\BibitemShut {NoStop}%
\bibitem [{\citenamefont {Zhai}\ and\ \citenamefont
  {Yao}(2020)}]{zhai2020Theory}%
  \BibitemOpen
  \bibfield  {author} {\bibinfo {author} {\bibfnamefont {D.}~\bibnamefont
  {Zhai}}\ and\ \bibinfo {author} {\bibfnamefont {W.}~\bibnamefont {Yao}},\
  }\href {https://link.aps.org/doi/10.1103/PhysRevMaterials.4.094002}
  {\bibfield  {journal} {\bibinfo  {journal} {Phys. Rev. Mater.}\ }\textbf
  {\bibinfo {volume} {4}},\ \bibinfo {pages} {094002} (\bibinfo {year}
  {2020})}\BibitemShut {NoStop}%
\bibitem [{\citenamefont {Fan}\ \emph {et~al.}(2024)\citenamefont {Fan},
  \citenamefont {Xiao},\ and\ \citenamefont {Yao}}]{fan2024Orbital}%
  \BibitemOpen
  \bibfield  {author} {\bibinfo {author} {\bibfnamefont {F.-R.}\ \bibnamefont
  {Fan}}, \bibinfo {author} {\bibfnamefont {C.}~\bibnamefont {Xiao}},\ and\
  \bibinfo {author} {\bibfnamefont {W.}~\bibnamefont {Yao}},\ }\href
  {https://doi.org/10.1103/PhysRevB.109.L041403} {\bibfield  {journal}
  {\bibinfo  {journal} {Phys. Rev. B}\ }\textbf {\bibinfo {volume} {109}},\
  \bibinfo {pages} {L041403} (\bibinfo {year} {2024})}\BibitemShut {NoStop}%
\bibitem [{\citenamefont {Jiang}\ \emph {et~al.}(2018)\citenamefont {Jiang},
  \citenamefont {Zhou}, \citenamefont {Dai},\ and\ \citenamefont
  {Wang}}]{jiang2018Antiferromagnetic}%
  \BibitemOpen
  \bibfield  {author} {\bibinfo {author} {\bibfnamefont {K.}~\bibnamefont
  {Jiang}}, \bibinfo {author} {\bibfnamefont {S.}~\bibnamefont {Zhou}},
  \bibinfo {author} {\bibfnamefont {X.}~\bibnamefont {Dai}},\ and\ \bibinfo
  {author} {\bibfnamefont {Z.}~\bibnamefont {Wang}},\ }\href
  {https://link.aps.org/doi/10.1103/PhysRevLett.120.157205} {\bibfield
  {journal} {\bibinfo  {journal} {Phys. Rev. Lett.}\ }\textbf {\bibinfo
  {volume} {120}},\ \bibinfo {pages} {157205} (\bibinfo {year}
  {2018})}\BibitemShut {NoStop}%
\bibitem [{\citenamefont {Liu}\ \emph {et~al.}(2024)\citenamefont {Liu},
  \citenamefont {He}, \citenamefont {Wang}, \citenamefont {Zhang},
  \citenamefont {Cao},\ and\ \citenamefont {Xiao}}]{liu2023Gatetunable}%
  \BibitemOpen
  \bibfield  {author} {\bibinfo {author} {\bibfnamefont {X.}~\bibnamefont
  {Liu}}, \bibinfo {author} {\bibfnamefont {Y.}~\bibnamefont {He}}, \bibinfo
  {author} {\bibfnamefont {C.}~\bibnamefont {Wang}}, \bibinfo {author}
  {\bibfnamefont {X.-W.}\ \bibnamefont {Zhang}}, \bibinfo {author}
  {\bibfnamefont {T.}~\bibnamefont {Cao}},\ and\ \bibinfo {author}
  {\bibfnamefont {D.}~\bibnamefont {Xiao}},\ }\href
  {https://journals.aps.org/prl/abstract/10.1103/PhysRevLett.132.146401}
  {\bibfield  {journal} {\bibinfo  {journal} {Phys. Rev. Lett.}\ }\textbf
  {\bibinfo {volume} {132}},\ \bibinfo {pages} {146401} (\bibinfo {year}
  {2024})}\BibitemShut {NoStop}%
\bibitem [{sup()}]{supp}%
  \BibitemOpen
  \href@noop {} {}\bibinfo {howpublished} {See Supplemental Material for the
  detailed derivations of Eq. (1), layer-projected Berry curvature, in-plane
  orbital magnetoelectric response, dipole Berry curvature polarizability, the
  details of Hartree-Fock mean-field method, dipole Berry curvature inversion
  at $\nu=-1$ and $-2$, interplay of interlayer tunneling and intralayer
  moir\'e potential, and protected scaling in the QAH phases, which includes
  Refs. \cite{fan2024Orbital, Ashvin2019,Qin2021, wu2019Topological,
  Xiao2022NLSOT, wang2023Fractional}}\BibitemShut {NoStop}%
\bibitem [{\citenamefont {Sun}\ \emph {et~al.}(2016)\citenamefont {Sun},
  \citenamefont {Zhang}, \citenamefont {Felser},\ and\ \citenamefont
  {Yan}}]{Sun2016}%
  \BibitemOpen
  \bibfield  {author} {\bibinfo {author} {\bibfnamefont {Y.}~\bibnamefont
  {Sun}}, \bibinfo {author} {\bibfnamefont {Y.}~\bibnamefont {Zhang}}, \bibinfo
  {author} {\bibfnamefont {C.}~\bibnamefont {Felser}},\ and\ \bibinfo {author}
  {\bibfnamefont {B.}~\bibnamefont {Yan}},\ }\href
  {https://doi.org/10.1103/PhysRevLett.117.146403} {\bibfield  {journal}
  {\bibinfo  {journal} {Phys. Rev. Lett.}\ }\textbf {\bibinfo {volume} {117}},\
  \bibinfo {pages} {146403} (\bibinfo {year} {2016})}\BibitemShut {NoStop}%
\bibitem [{\citenamefont {Xiao}\ and\ \citenamefont {Niu}(2021)}]{Xiao2021CS}%
  \BibitemOpen
  \bibfield  {author} {\bibinfo {author} {\bibfnamefont {C.}~\bibnamefont
  {Xiao}}\ and\ \bibinfo {author} {\bibfnamefont {Q.}~\bibnamefont {Niu}},\
  }\href {https://doi.org/10.1103/PhysRevB.104.L241411} {\bibfield  {journal}
  {\bibinfo  {journal} {Phys. Rev. B}\ }\textbf {\bibinfo {volume} {104}},\
  \bibinfo {pages} {L241411} (\bibinfo {year} {2021})}\BibitemShut {NoStop}%
\bibitem [{\citenamefont {Lee}\ \emph {et~al.}(2019)\citenamefont {Lee},
  \citenamefont {Khalaf}, \citenamefont {Liu}, \citenamefont {Liu},
  \citenamefont {Hao}, \citenamefont {Kim},\ and\ \citenamefont
  {Vishwanath}}]{Ashvin2019}%
  \BibitemOpen
  \bibfield  {author} {\bibinfo {author} {\bibfnamefont {J.~Y.}\ \bibnamefont
  {Lee}}, \bibinfo {author} {\bibfnamefont {E.}~\bibnamefont {Khalaf}},
  \bibinfo {author} {\bibfnamefont {S.}~\bibnamefont {Liu}}, \bibinfo {author}
  {\bibfnamefont {X.}~\bibnamefont {Liu}}, \bibinfo {author} {\bibfnamefont
  {Z.}~\bibnamefont {Hao}}, \bibinfo {author} {\bibfnamefont {P.}~\bibnamefont
  {Kim}},\ and\ \bibinfo {author} {\bibfnamefont {A.}~\bibnamefont
  {Vishwanath}},\ }\href {https://doi.org/10.1038/s41467-019-12981-1}
  {\bibfield  {journal} {\bibinfo  {journal} {Nat. Commun.}\ }\textbf {\bibinfo
  {volume} {10}},\ \bibinfo {pages} {5333} (\bibinfo {year}
  {2019})}\BibitemShut {NoStop}%
\bibitem [{\citenamefont {Qin}\ and\ \citenamefont
  {MacDonald}(2021)}]{Qin2021}%
  \BibitemOpen
  \bibfield  {author} {\bibinfo {author} {\bibfnamefont {W.}~\bibnamefont
  {Qin}}\ and\ \bibinfo {author} {\bibfnamefont {A.~H.}\ \bibnamefont
  {MacDonald}},\ }\href {https://doi.org/10.1103/PhysRevLett.127.097001}
  {\bibfield  {journal} {\bibinfo  {journal} {Phys. Rev. Lett.}\ }\textbf
  {\bibinfo {volume} {127}},\ \bibinfo {pages} {097001} (\bibinfo {year}
  {2021})}\BibitemShut {NoStop}%
\bibitem [{\citenamefont {Culcer}\ \emph {et~al.}(2004)\citenamefont {Culcer},
  \citenamefont {Sinova}, \citenamefont {Sinitsyn}, \citenamefont {Jungwirth},
  \citenamefont {MacDonald},\ and\ \citenamefont {Niu}}]{Dimi2004}%
  \BibitemOpen
  \bibfield  {author} {\bibinfo {author} {\bibfnamefont {D.}~\bibnamefont
  {Culcer}}, \bibinfo {author} {\bibfnamefont {J.}~\bibnamefont {Sinova}},
  \bibinfo {author} {\bibfnamefont {N.~A.}\ \bibnamefont {Sinitsyn}}, \bibinfo
  {author} {\bibfnamefont {T.}~\bibnamefont {Jungwirth}}, \bibinfo {author}
  {\bibfnamefont {A.~H.}\ \bibnamefont {MacDonald}},\ and\ \bibinfo {author}
  {\bibfnamefont {Q.}~\bibnamefont {Niu}},\ }\href
  {https://doi.org/10.1103/PhysRevLett.93.046602} {\bibfield  {journal}
  {\bibinfo  {journal} {Phys. Rev. Lett.}\ }\textbf {\bibinfo {volume} {93}},\
  \bibinfo {pages} {046602} (\bibinfo {year} {2004})}\BibitemShut {NoStop}%
\bibitem [{\citenamefont {Ma}\ \emph {et~al.}(2019)\citenamefont {Ma},
  \citenamefont {Xu}, \citenamefont {Shen}, \citenamefont {MacNeill},
  \citenamefont {Fatemi}, \citenamefont {Chang}, \citenamefont {Mier~Valdivia},
  \citenamefont {Wu}, \citenamefont {Du}, \citenamefont {Hsu}, \citenamefont
  {Fang}, \citenamefont {Gibson}, \citenamefont {Watanabe}, \citenamefont
  {Taniguchi}, \citenamefont {Cava}, \citenamefont {Kaxiras}, \citenamefont
  {Lu}, \citenamefont {Lin}, \citenamefont {Fu}, \citenamefont {Gedik},\ and\
  \citenamefont {Jarillo-Herrero}}]{Ma2019}%
  \BibitemOpen
  \bibfield  {author} {\bibinfo {author} {\bibfnamefont {Q.}~\bibnamefont
  {Ma}}, \bibinfo {author} {\bibfnamefont {S.-Y.}\ \bibnamefont {Xu}}, \bibinfo
  {author} {\bibfnamefont {H.}~\bibnamefont {Shen}}, \bibinfo {author}
  {\bibfnamefont {D.}~\bibnamefont {MacNeill}}, \bibinfo {author}
  {\bibfnamefont {V.}~\bibnamefont {Fatemi}}, \bibinfo {author} {\bibfnamefont
  {T.-R.}\ \bibnamefont {Chang}}, \bibinfo {author} {\bibfnamefont {A.~M.}\
  \bibnamefont {Mier~Valdivia}}, \bibinfo {author} {\bibfnamefont
  {S.}~\bibnamefont {Wu}}, \bibinfo {author} {\bibfnamefont {Z.}~\bibnamefont
  {Du}}, \bibinfo {author} {\bibfnamefont {C.-H.}\ \bibnamefont {Hsu}},
  \bibinfo {author} {\bibfnamefont {S.}~\bibnamefont {Fang}}, \bibinfo {author}
  {\bibfnamefont {Q.~D.}\ \bibnamefont {Gibson}}, \bibinfo {author}
  {\bibfnamefont {K.}~\bibnamefont {Watanabe}}, \bibinfo {author}
  {\bibfnamefont {T.}~\bibnamefont {Taniguchi}}, \bibinfo {author}
  {\bibfnamefont {R.~J.}\ \bibnamefont {Cava}}, \bibinfo {author}
  {\bibfnamefont {E.}~\bibnamefont {Kaxiras}}, \bibinfo {author} {\bibfnamefont
  {H.-Z.}\ \bibnamefont {Lu}}, \bibinfo {author} {\bibfnamefont
  {H.}~\bibnamefont {Lin}}, \bibinfo {author} {\bibfnamefont {L.}~\bibnamefont
  {Fu}}, \bibinfo {author} {\bibfnamefont {N.}~\bibnamefont {Gedik}},\ and\
  \bibinfo {author} {\bibfnamefont {P.}~\bibnamefont {Jarillo-Herrero}},\
  }\href {https://doi.org/10.1038/s41586-018-0807-6} {\bibfield  {journal}
  {\bibinfo  {journal} {Nature}\ }\textbf {\bibinfo {volume} {565}},\ \bibinfo
  {pages} {337} (\bibinfo {year} {2019})}\BibitemShut {NoStop}%
\bibitem [{\citenamefont {Kang}\ \emph {et~al.}(2019)\citenamefont {Kang},
  \citenamefont {Li}, \citenamefont {Sohn}, \citenamefont {Shan},\ and\
  \citenamefont {Mak}}]{Kang2019}%
  \BibitemOpen
  \bibfield  {author} {\bibinfo {author} {\bibfnamefont {K.}~\bibnamefont
  {Kang}}, \bibinfo {author} {\bibfnamefont {T.}~\bibnamefont {Li}}, \bibinfo
  {author} {\bibfnamefont {E.}~\bibnamefont {Sohn}}, \bibinfo {author}
  {\bibfnamefont {J.}~\bibnamefont {Shan}},\ and\ \bibinfo {author}
  {\bibfnamefont {K.~F.}\ \bibnamefont {Mak}},\ }\href
  {https://doi.org/10.1038/s41563-019-0294-7} {\bibfield  {journal} {\bibinfo
  {journal} {Nat. Mater.}\ }\textbf {\bibinfo {volume} {18}},\ \bibinfo {pages}
  {324} (\bibinfo {year} {2019})}\BibitemShut {NoStop}%
\bibitem [{\citenamefont {Du}\ \emph {et~al.}(2021)\citenamefont {Du},
  \citenamefont {Lu},\ and\ \citenamefont {Xie}}]{Lu2021}%
  \BibitemOpen
  \bibfield  {author} {\bibinfo {author} {\bibfnamefont {Z.~Z.}\ \bibnamefont
  {Du}}, \bibinfo {author} {\bibfnamefont {H.-Z.}\ \bibnamefont {Lu}},\ and\
  \bibinfo {author} {\bibfnamefont {X.}~\bibnamefont {Xie}},\ }\href
  {https://doi.org/10.1038/s42254-021-00359-6} {\bibfield  {journal} {\bibinfo
  {journal} {Nat. Rev. Phys.}\ }\textbf {\bibinfo {volume} {3}},\ \bibinfo
  {pages} {744} (\bibinfo {year} {2021})}\BibitemShut {NoStop}%
\bibitem [{\citenamefont {Dong}\ \emph {et~al.}(2015)\citenamefont {Dong},
  \citenamefont {Liu}, \citenamefont {Cheong},\ and\ \citenamefont
  {Ren}}]{dong2015multiferroic}%
  \BibitemOpen
  \bibfield  {author} {\bibinfo {author} {\bibfnamefont {S.}~\bibnamefont
  {Dong}}, \bibinfo {author} {\bibfnamefont {J.-M.}\ \bibnamefont {Liu}},
  \bibinfo {author} {\bibfnamefont {S.-W.}\ \bibnamefont {Cheong}},\ and\
  \bibinfo {author} {\bibfnamefont {Z.}~\bibnamefont {Ren}},\ }\href
  {https://doi.org/10.1080/00018732.2015.1114338} {\bibfield  {journal}
  {\bibinfo  {journal} {Adv. Phys.}\ }\textbf {\bibinfo {volume} {64}},\
  \bibinfo {pages} {519} (\bibinfo {year} {2015})}\BibitemShut {NoStop}%
\bibitem [{\citenamefont {Spaldin}\ and\ \citenamefont
  {Ramesh}(2019)}]{spaldin2019advances}%
  \BibitemOpen
  \bibfield  {author} {\bibinfo {author} {\bibfnamefont {N.~A.}\ \bibnamefont
  {Spaldin}}\ and\ \bibinfo {author} {\bibfnamefont {R.}~\bibnamefont
  {Ramesh}},\ }\href {https://doi.org/10.1038/s41563-018-0275-2} {\bibfield
  {journal} {\bibinfo  {journal} {Nat. Mater.}\ }\textbf {\bibinfo {volume}
  {18}},\ \bibinfo {pages} {203} (\bibinfo {year} {2019})}\BibitemShut
  {NoStop}%
\bibitem [{\citenamefont {Stauber}\ \emph {et~al.}(2018)\citenamefont
  {Stauber}, \citenamefont {Low},\ and\ \citenamefont
  {G\'omez-Santos}}]{Stauber2018}%
  \BibitemOpen
  \bibfield  {author} {\bibinfo {author} {\bibfnamefont {T.}~\bibnamefont
  {Stauber}}, \bibinfo {author} {\bibfnamefont {T.}~\bibnamefont {Low}},\ and\
  \bibinfo {author} {\bibfnamefont {G.}~\bibnamefont {G\'omez-Santos}},\ }\href
  {https://doi.org/10.1103/PhysRevLett.120.046801} {\bibfield  {journal}
  {\bibinfo  {journal} {Phys. Rev. Lett.}\ }\textbf {\bibinfo {volume} {120}},\
  \bibinfo {pages} {046801} (\bibinfo {year} {2018})}\BibitemShut {NoStop}%
\bibitem [{\citenamefont {Zhai}\ \emph {et~al.}(2023)\citenamefont {Zhai},
  \citenamefont {Chen}, \citenamefont {Xiao},\ and\ \citenamefont
  {Yao}}]{zhai2023Timereversal}%
  \BibitemOpen
  \bibfield  {author} {\bibinfo {author} {\bibfnamefont {D.}~\bibnamefont
  {Zhai}}, \bibinfo {author} {\bibfnamefont {C.}~\bibnamefont {Chen}}, \bibinfo
  {author} {\bibfnamefont {C.}~\bibnamefont {Xiao}},\ and\ \bibinfo {author}
  {\bibfnamefont {W.}~\bibnamefont {Yao}},\ }\href
  {https://www.nature.com/articles/s41467-023-37644-0} {\bibfield  {journal}
  {\bibinfo  {journal} {Nat. Commun.}\ }\textbf {\bibinfo {volume} {14}},\
  \bibinfo {pages} {1961} (\bibinfo {year} {2023})}\BibitemShut {NoStop}%
\bibitem [{\citenamefont {Bauer}\ \emph {et~al.}(2022)\citenamefont {Bauer},
  \citenamefont {Tang}, \citenamefont {Iguchi},\ and\ \citenamefont
  {Uchida}}]{bauer2022magnonics}%
  \BibitemOpen
  \bibfield  {author} {\bibinfo {author} {\bibfnamefont {G.~E.}\ \bibnamefont
  {Bauer}}, \bibinfo {author} {\bibfnamefont {P.}~\bibnamefont {Tang}},
  \bibinfo {author} {\bibfnamefont {R.}~\bibnamefont {Iguchi}},\ and\ \bibinfo
  {author} {\bibfnamefont {K.-i.}\ \bibnamefont {Uchida}},\ }\href
  {https://www.sciencedirect.com/science/article/abs/pii/S0304885321007344}
  {\bibfield  {journal} {\bibinfo  {journal} {J. Magn. Magn. Mater.}\ }\textbf
  {\bibinfo {volume} {541}},\ \bibinfo {pages} {168468} (\bibinfo {year}
  {2022})}\BibitemShut {NoStop}%
\bibitem [{\citenamefont {Xiao}\ \emph {et~al.}(2022)\citenamefont {Xiao},
  \citenamefont {Liu}, \citenamefont {Wu}, \citenamefont {Wang}, \citenamefont
  {Niu},\ and\ \citenamefont {Yang}}]{Xiao2022NLSOT}%
  \BibitemOpen
  \bibfield  {author} {\bibinfo {author} {\bibfnamefont {C.}~\bibnamefont
  {Xiao}}, \bibinfo {author} {\bibfnamefont {H.}~\bibnamefont {Liu}}, \bibinfo
  {author} {\bibfnamefont {W.}~\bibnamefont {Wu}}, \bibinfo {author}
  {\bibfnamefont {H.}~\bibnamefont {Wang}}, \bibinfo {author} {\bibfnamefont
  {Q.}~\bibnamefont {Niu}},\ and\ \bibinfo {author} {\bibfnamefont {S.~A.}\
  \bibnamefont {Yang}},\ }\href
  {https://doi.org/10.1103/PhysRevLett.129.086602} {\bibfield  {journal}
  {\bibinfo  {journal} {Phys. Rev. Lett.}\ }\textbf {\bibinfo {volume} {129}},\
  \bibinfo {pages} {086602} (\bibinfo {year} {2022})}\BibitemShut {NoStop}%
\bibitem [{\citenamefont {Wang}\ \emph {et~al.}(2024)\citenamefont {Wang},
  \citenamefont {Zhang}, \citenamefont {Liu}, \citenamefont {He}, \citenamefont
  {Xu}, \citenamefont {Ran}, \citenamefont {Cao},\ and\ \citenamefont
  {Xiao}}]{wang2023Fractional}%
  \BibitemOpen
  \bibfield  {author} {\bibinfo {author} {\bibfnamefont {C.}~\bibnamefont
  {Wang}}, \bibinfo {author} {\bibfnamefont {X.-W.}\ \bibnamefont {Zhang}},
  \bibinfo {author} {\bibfnamefont {X.}~\bibnamefont {Liu}}, \bibinfo {author}
  {\bibfnamefont {Y.}~\bibnamefont {He}}, \bibinfo {author} {\bibfnamefont
  {X.}~\bibnamefont {Xu}}, \bibinfo {author} {\bibfnamefont {Y.}~\bibnamefont
  {Ran}}, \bibinfo {author} {\bibfnamefont {T.}~\bibnamefont {Cao}},\ and\
  \bibinfo {author} {\bibfnamefont {D.}~\bibnamefont {Xiao}},\ }\href
  {https://link.aps.org/doi/10.1103/PhysRevLett.132.036501} {\bibfield
  {journal} {\bibinfo  {journal} {Phys. Rev. Lett.}\ }\textbf {\bibinfo
  {volume} {132}},\ \bibinfo {pages} {036501} (\bibinfo {year}
  {2024})}\BibitemShut {NoStop}%
\end{thebibliography}%

\end{document}